\documentclass[aps,pre,twocolumn]{revtex4}

\usepackage{graphicx,amssymb,amsmath,times}

\begin{document}

\title{Modeling bursts and heavy tails in human dynamics}

\author{Alexei V\'azquez$^{1,2}$, Jo\~ao Gama Oliveira$^{2,3}$, Zolt\'an 
Dezs\H{o}$^2$, Kwang-Il Goh$^{1,2}$, Imre Kondor$^4$ and Albert-L\'aszl\'o 
Barab\'asi$^{1,2}$}

\affiliation{$^1$Center for Cancer System Biology, Dana Farber Cancer 
Institute, Harvard Medical School, 44 Binney St, Boston, MA 02115, USA}

\affiliation{$^2$Department of Physics and Center for Complex Networks
Research, University of Notre Dame, IN 46556, USA}

\affiliation{$^3$Departamento de F\'{i}sica, Universidade de Aveiro, 
Campus Universit\'ario de Santiago, 3810-193 Aveiro, Portugal} 

\affiliation{$^4$Collegium Budapest, Szenth\'aroms\'ag u. 2, H-1014 
Budapest, Hungary}

\date{\today}

\begin{abstract}

The dynamics of many social, technological and economic phenomena are
driven by individual human actions, turning the quantitative understanding
of human behavior into a central question of modern science. Current
models of human dynamics, used from risk assessment to communications,
assume that human actions are randomly distributed in time and thus well
approximated by Poisson processes. Here we provide direct evidence that
for five human activity patterns, such as email and letter based
communications, web browsing, library visits and stock trading,
the timing of individual human actions follow non-Poisson statistics,
characterized by bursts of rapidly occurring events separated by long
periods of inactivity. We show that the bursty nature of human behavior is
a consequence of a decision based queuing process: when individuals
execute tasks based on some perceived priority, the timing of the tasks
will be heavy tailed, most tasks being rapidly executed, while a few
experiencing very long waiting times. In contrast, priority blind
execution is well approximated by uniform interevent statistics. We
discuss two queueing models that capture human activity. The first model
assumes that there are no limitations on the number of tasks an individual
can hadle at any time, predicting that the waiting time of the individual
tasks follow a heavy tailed distribution $P(\tau_w)\sim\tau_w^{-\alpha}$
with $\alpha=3/2$. The second model imposes limitations on the queue
length, resulting in a heavy tailed waiting time distribution
characterized by $\alpha=1$. We provide empirical evidence supporting the
relevance of these two models to human activity patterns, showing that
while emails, web browsing and library visitation display $\alpha=1$, the
surface mail based communication belongs to the $\alpha=3/2$ universality
class. Finally, we discuss possible extension of the proposed queueing
models and outline some future challenges in exploring the statistical
mechanisms of human dynamics. These findings have important implications
not only for our quantitative understanding of human activity patterns,
but also for resource management and service allocation in both
communications and retail.

\end{abstract}


\maketitle


\section{Introduction}
\label{sec:intro}

Humans participate on a daily basis in a large number of distinct
activities, from electronic communication, such as sending emails
or browsing the web, to initiating financial transactions or engaging in
entertainment and sports. Given the number of factors that determine the
timing of each action, ranging from work and sleep patterns to resource
availability, it appears impossible to seek regularities in the apparently
random human activity patterns, apart from the obvious daily and seasonal
periodicities. Therefore, in contrast with the accurate predictive tools
common in physical sciences, forecasting human and social patterns remains
a difficult and often elusive goal. Yet, the need to understand the timing
of human actions is increasingly important. Indeed, uncovering the laws
governing human dynamics in a quantitative manner is of major scientific
interest, requiring us to address the factors that determine the timing of
human actions. But these questions are driven by applications as well:
most human actions have a strong impact on resource allocation, from phone
line availability and bandwidth allocation in the case of Internet or Web
use, all the way to the design of physical space for retail or service
oriented institutions. Despite these fundamental and practical driving
forces, our understanding of the timing of human initiated actions is
rather limited at present.

To be sure, the interest in addressing the timing of events in human
dynamics is not new: it has a long history in the mathematical literature,
leading to the development of some of the key concepts in probability
theory \cite{fellerII}, and has reemerged at the beginning of the
$20^{\text{th}}$ century as the design problems surrounding the phone
system required a quantitative understanding of the call patterns of
individuals. But most current models of human activity assume that human
actions are performed at constant rate, meaning that a user has a fixed
probability to engage in a specific action within a given time interval.
These models approximate the timing of human actions with a Poisson
process, in which the time interval between two consecutive actions by the
same individual, called the waiting or inter-event time, follows an
exponential distribution \cite{Haight67}. Poisson processes are at the
heart of the celebrated Erlang formula \cite{erlang}, predicting the
number of phone lines required in an institution, and they represent the
basic approximation in the design of most currently used Internet
protocols and routers \cite{phone-design}. Yet, the availability of large
datasets recording selected human activity patterns increasingly question
the validity of the Poisson approximation. Indeed, an increasing number of
recent measurements indicate that the timing of many human actions
systematically deviate from the Poisson prediction, the waiting or
inter-event times being better approximated by a heavy tailed or Pareto
distribution \cite{barabasi05,joao05,zoli,vazquez05}. The difference
between a Poisson and a heavy tailed behavior is striking: the exponential
decay of a Poisson distribution forces the consecutive events to follow
each other at relatively regular time intervals and forbids very long
waiting times. In contrast, the slowly decaying heavy tailed processes
allow for very long periods of inactivity that separate bursts of
intensive activity.

We have recently proposed that the bursty nature of human dynamics is a
consequence of a queuing process driven by human decision making
\cite{barabasi05}: whenever an individual is presented with multiple tasks
and chooses among them based on some perceived priority parameter, the
waiting time of the various tasks will be Pareto distributed. In contrast,
first-come-first-serve and random task execution, common in most service
oriented or computer driven environments, lead to a uniform Poisson-like
dynamics. Yet, this work has generated just as many questions as it
resolved. What are the different classes of processes that are relevant
for human dynamics? What determines the scaling exponents? Do we have
discrete universality classes (and if so how many) as in critical
phenomena \cite{stanley}, or the exponents characterizing the heavy tails
can take up arbitrary values, as it is the case in network theory
\cite{review1,review2,vespignani-book}? Is human dynamics always heavy
tailed?

In this paper we aim to address some of these questions by studying the
different universality classes that can appear as a result of the queuing
of human activities. We first review, in Sect. \ref{sec:poisson}, the
frequently used Poisson approximation, which predicts an exponential
distribution of interevent times. In Sect. \ref{sec:empirical} we present
evidence that the interevent time probability density function (pdf)
$P(\tau)$ of many human activities is characterized by the power law tail

\begin{equation}
P(\tau)\sim\tau^{-\alpha}\ . 
\label{E:eq1}
\end{equation}

\noindent In Sect. \ref{sec:model} we discuss the general characteristics
of the queueing models that govern how humans time their various
activities. In Sects. \ref{sec:cobham}-\ref{sec:barabasi} we study two
classes of queuing models designed to capture human activity patterns. We
find that restrictions on the queue length play an important role in
determining the scaling of the queuing process, allowing us to document
the existence of two distinct universality classes, one characterized by
$\alpha=3/2$ (Sect. \ref{sec:cobham}) and the other by $\alpha=1$ (Sect.
\ref{sec:barabasi}). In Sect. \ref{sec:wait_interevent} we discuss the
relationship between interevent and waiting times. Finally, in Sec.
\ref{sec:discussion} we discuss the applicability of these models to
explain the empirical data, as well as outline future challenges in
modeling human dynamics.

\section{Poisson processes}
\label{sec:poisson}

Consider an activity performed with some regularity, such as sending
emails, placing phone calls, visiting a library, or browsing the web. We
can keep track of this activity by recording the timing of each event, for
example the time each email is sent by an individual. The time between two
consecutive events we call {\it the interevent time} for the monitored
activity and will be denoted by $\tau$. Given that the interevent time can
be explicitly measured for selected activities, it serves as a test of our
ability to understand and model human dynamics: proper models should be
able to capture its statistical properties.

\begin{figure}
\centerline{\includegraphics[width=3.2in]{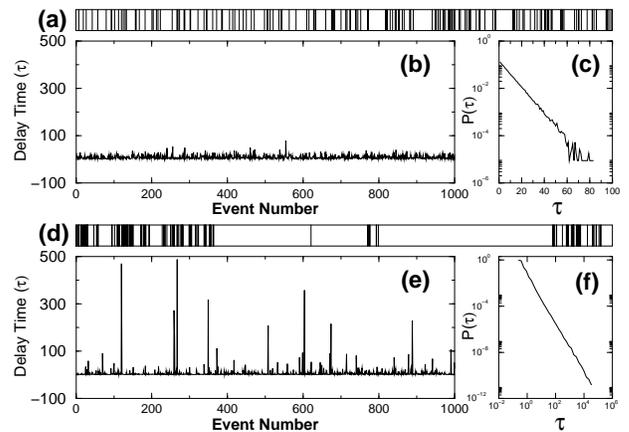}}

\caption{The difference between the activity patterns predicted by a
Poisson process (top) and the heavy tailed distributions observed in human
dynamics (bottom). {\bf (a)} Succession of events predicted by a Poisson
process, which assumes that in any moment events take place with
probability $q$. The horizontal axis denotes time, each vertical line
corresponding to an individual event. Note that the interevent times are
comparable to each other, long delays being virtually absent. {\bf (b)}
The absence of long delays is visible on the plot showing the delay times
$\tau$ for 1,000 consecutive events, the size of each vertical line
corresponding to the gaps seen in (a).  {\bf (c)} The probability of
finding exactly $n$ events within a fixed time interval is $ {\cal
P}(n;q)= e^{-q t} (q t)^n/n!$, which predicts that for a Poisson process
the inter-event time distribution follows $P(\tau)=q e^{-q\tau}$, shown on
a log-linear plot in (c) for the events displayed in (a, b). {\bf (d)} The
succession of events for a heavy tailed distribution. {\bf (e)} The
waiting time $\tau$ of 1,000 consecutive events, where the mean event time
was chosen to coincide with the mean event time of the Poisson process
shown in (a-c). Note the large spikes in the plot, corresponding to very
long delay times.  (b) and (e) have the same vertical scale, allowing to
compare the regularity of a Poisson process with the bursty nature of the
heavy tailed process.  {\bf (f)} Delay time distribution $P(\tau) \simeq
\tau^{-2}$ for the heavy tailed process shown in (d,e), appearing as a
straight line with slope -2 on a log-log plot.  The signal shown in (d-f)
was generated using $\gamma=1$ in the stochastic priority list model
discussed in Appendix \ref{sec:preferential}.}

\label{fig2}
\end{figure}

The most primitive model of human activity would assume that human actions
are fundamentally periodic, with a period determined by the daily sleep
patterns. Yet, while certain periodicity is certainly present, the timing
of most human actions are highly stochastic. Indeed, periodic models are
hopeless in capturing the time we check out a book from the library,
beyond telling us that it should be within the library's operation hours.
The first and still most widely used stochastic model of human activity
assumes that the tasks are executed independently from each other at a
constant rate $\lambda$, so that the time resolved activity of an
individual is well approximated by a Poisson process \cite{Haight67}. In
this case the probability density function (pdf) of the recorded
interevent times has the exponential form

\begin{equation}
P(\tau) = \lambda e^{-\lambda \tau}\ .
\label{PPoisson}
\end{equation}

\noindent In practice this means that the predicted activity pattern,
while stochastic, will display some regularity in time, events following
each other on average at $\tau\approx\langle\tau\rangle = 1/\lambda$
intervals. Indeed, given that for a Poisson process $\sigma = \sqrt{
\langle\tau^2\rangle - \langle\tau\rangle^2} = \langle\tau\rangle$ is
finite, very long waiting times ({\it i.e.} large temporal gaps in the
sequence of events) are exponentially rare. This is illustrated in Fig.
\ref{fig2}a, where we show a sequence of events generated by a Poisson
process, appearing uniformly distributed in time (but not periodic).

The Poisson process was originally introduced by Poisson in his major work
applying probability concepts to the administration of justice
\cite{poisson}. Today it is widely used to quantify the consequences of
human actions, such as modeling traffic flow patterns or accident
frequencies \cite{Haight67}, and is commercially used in call center
staffing \cite{call-center}, inventory control \cite{inventory}, or to
estimate the number of congestion caused blocked calls in mobile
communications \cite{phone-design}. It has been established as a basic
model of human activity patterns at a time when data collection
capabilities on human behavior were rather limited. In the past few years,
however, thanks to detailed computer based data collection methods, there
is increasing evidence that the Poisson approximation fails to capture the
timing of many human actions.

\begin{figure}
\centerline{\includegraphics[width=3.2in]{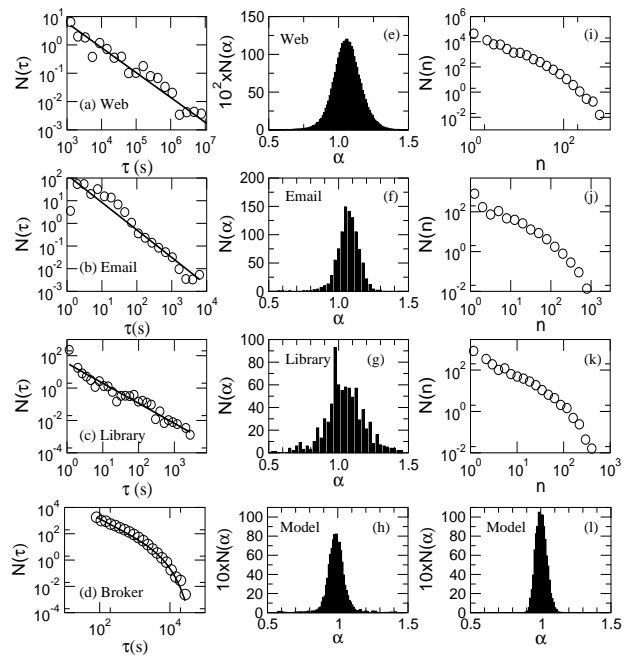}}

\caption{ (a) The interevent time distribution between (a) two consecutive
visits of a webportal by a single user; (b) two consecutive library loans
made by a single individual; (c) two consecutive emails sent out by a
user. For (a-c) we show as a straight line the $\alpha=1$ scaling. (d) The
interevent time distribution between two consecutive transactions made by
a stock broker. The distribution follows a power-law with the exponential
cut-off $P(\tau) \sim \tau^{-1.3}\exp(-\tau/\tau_0)$.  (e-g) The
distribution of the exponents ($\alpha$) characterizing the interevent
time distribution of users browsing the web portal (e), individual loans
from the library (f) and the emails sent by different individuals (g). The
exponent $\alpha$ was determined only for users whose total activity
levels exceeded certain thresholds, the values used being 15 web visits
(e), 15 emails (f) and 10 books (g). (h,l) We numerically generate for
10,000 individuals interevent time distributions following a power-law
with exponent $\alpha=1$. The distribution of the measured exponents
follows a normal distribution similar to the distribution observed in
(e-g). If we double the time window of the simulation (h) the deviation
around the average becomes much smaller (l). (i-k) The distribution of the
number of events in the studied systems: number of HTML hits for each user
(i), the number of books checked out by each user (j) and the number of
emails sent by different individuals (k), indicating that the overall
activity patterns of individuals is also heavy tailed. }

\label{F:fig2}
\end{figure}

\section{Empirical results}
\label{sec:empirical}

Evidence that non-Poisson activity patterns characterize human activity
has first emerged in computer communications, where the timing of many
human driven events is automatically recorded. For example, measurements
capturing the distribution of the time differences between consecutive
instant messages sent by individuals during online chats \cite{Instant}
have found evidence of heavy tailed statistics. Professional tasks, such
as the timing of job submissions on a supercomputer \cite{supercomputers},
directory listings and file transfers (FTP requests) initiated by
individual users \cite{ftp} were also reported to display non-Poisson
features. Similar patterns emerge in economic transactions
\cite{mainardi00,mainardi02}, in the number of hourly trades in a given
security \cite{economic1} or the time interval distribution between
individual trades in currency futures \cite{economic2}. Finally, heavy
tailed distributions characterize entertainment related events, such as
the time intervals between consecutive online games played by users
\cite{games}. Note, however, that while these datasets provide clear
evidence for non-Poisson human activity patterns, most of them do not
resolve individual human behavior, but capture only the aggregated
behavior of a large number of users. For example, the dataset recording
the timing of the job submissions looks at the timing of {\it all jobs}
submitted to a computer, by any user. Thus for these measurements the
interevent time does not characterize a {\it single} user but rather a
{\it population} of users. Given the extensive evidence that the activity
distribution of the individuals in a population is heavy tailed, these
measurements have difficulty capturing the origin of the observed heavy
tailed patterns. For example, while most people send only a few emails per
day, a few send a very large number on a daily basis \cite{eckmann,ebel}.

If the activity pattern of a large number of users is simultaneously
captured, it is not clear where the observed heavy tails come from: are
they rooted in the activity of a single individual, or rather in the heavy
tailed distribution of user activities? Therefore, when it comes to our
quest to understand human dynamics, datasets that capture the long term
activity pattern of a {\it single} individual are of particular value. To
our best knowledge only three papers have taken this approach, capturing
the timing of printing jobs submitted by users \cite{Maya}, the email
activity patterns of individual email users \cite{barabasi05,eckmann} and
the browsing pattern of users visiting a major web portal \cite{zoli}.
These measurements offer direct evidence that the heavy tailed activity
patterns emerge at the level of a {\it single} individual, and are not a
consequence of the heterogeneous distribution of user activity. Despite
this evidence, a number of questions remain unresolved: Is there a single
scaling exponent characterizing all users, or rather each user has its own
exponent? What is the range of these exponents? Next we aim to address
these questions through the study of six datasets, each capturing
individual human activity patterns of different nature. First we describe
the datasets and the collection methods, followed by a quantitative
characterization of the observed human activity patterns.

{\it Web browsing:} Automatically assigned cookies allow us to reconstruct
the browsing history of approximately 250,000 unique visitors of the
largest Hungarian news and entertainment portal (origo.hu), which provides
online news and magazines, community pages, software downloads, free email
and search engine, capturing 40\% of all internal Web traffic in Hungary
\cite{zoli,racz04}. The portal receives 6,500,000 HTML hits on a typical
workday. We used the log files of the portal to collect the visitation
pattern of each visitor between 11/08/02 and 12/08/02, recording with
second resolution the timing of each download by each visitor \cite{zoli}.  
The interevent time, $\tau$, was defined as the time interval between
consecutive page downloads (clicks) by the same visitor.

{\it Email activity patterns:} This dataset contains the email
exchange between individuals in a university environment, capturing
the sender, recipient and the time of each email sent during a three
and six month period by 3,188 \cite{eckmann} and 9,665 \cite{ebel}
users, respectively. We focused here on the data collected by
Eckmann \cite{eckmann}, which records 129,135 emails with second
resolution. The interevent time corresponds to the time between two
consecutive emails sent by the same user.

{\it Library loans:} The data contains the time with second
resolution at which books or periodicals were checked out from the
library by the faculty at University of Notre Dame during a three
year period. The number of unique individuals in the dataset is
2,247, together participating in a total of 48,409 transactions. The
interevent time corresponds to the time difference between
consecutive books or periodicals checked out by the same patron.

{\it Trade transactions:} A dataset recording all transactions (buy/sell)
initiated by a stock broker at a Central European bank between 06/99 and
5/03 helps us quantify the professional activity of a single individual,
giving a glimpse on the human activity patterns driving economic
phenomena. In a typical day the first transactions start at 7AM and end at
7PM and the average number of transactions initiated by the dealer in one
day is around 10, resulting in a total of 54,374 transactions. The
interevent time represents the time between two consecutive transactions
by the broker. The gap between the last transaction at the end of one day
and the first transaction at the beginning of the next trading day was
ignored.

{\it The correspondence patterns of Einstein, Darwin and Freud:} We start
from a record containing the sender, recipient and the date of each letter
\cite{einstein,darwin,freud} sent or received by the three scientists
during their lifetime. The databases used in our study were provided by
the Darwin Correspondence Project
(http://www.lib.cam.ac.uk/Departments/Darwin/), the Einstein Papers
Project (http://www.einstein.caltech.edu/) and the Freud Museum of London
(http://www.freud.org.uk). Each dataset contains the information about
each sent/received letter in the following format:  SENDER, RECIPIENT,
DATE, where either the sender or the recipient is Einstein, Darwin or
Freud. The Darwin dataset contained a record of a total of 7,591 letters
sent and 6,530 letters received by Darwin (a total of 14,121 letters).
Similarly, the Einstein database contained 14,512 letters sent and 16,289
letters received (total of 30,801). For Freud we have 3183 (2675) sent
(received) leters. Note that 1,541 letters in the Darwin database and
1,861 letters in the Einstein database were not dated or were assigned
only potential time intervals spanning days or months. We discarded these
letters from the dataset.  Furthermore, the dataset is naturally
incomplete, as not all letters written or received by these scientists
were preserved. Yet, assuming that letters are lost at a uniform rate,
they should not affect our main findings. For these three datasets we do
not focus on the interevent times, but rather the {\it response} or {\it
waiting times} $\tau_w$. The waiting time, $\tau_w$, represents the time
interval between the date of a letter received from a given person, and
the date of the next letter from Darwin, Einstein or Freud to him or her,
{\it i.e.} the time the letter waited on their desk before a response is
being sent. To analyze Einstein, Darwin, and Freud's response time we have
followed the following procedure: if individual A sent a letter to
Einstein on DATE1, we search for the next letter from Einstein to
individual A, sent on DATE2, the response time representing the time
difference $\tau={\rm DATE2-DATE1}$, expressed in days. If there are
multiple letters from Einstein to the recipient, we always consider the
first letter as the response, and discard the later ones. Missing letters
could increase the response time, the magnitude of this effect depending
on the overall frequency of communication between the respective
correspondence partners. Yet, if the response time follows a distribution
with an exponential tail, then randomly distributed missing letters would
not generate a power law waiting time: they would only shift shift the
exponential waiting times to longer average values. Thus the observed
power law cannot be attributed to data incompleteness.

In the following we will break our discussion in three subsections, each 
focusing on a specific class of behavior observed in the studied 
individual activity patterns.

\subsection{The $\alpha=1$ universality class: Web browsing, email, and 
library datasets}
\label{sec:alpha1}

In Fig. \ref{F:fig2}a-c we show the interevent time distribution between
consecutive events for a single individual for the first four studied
databases: Web browsing, email, and library visitation. For these
datasets we find that the interevent time distribution has a power-law
tail

\begin{equation}
P(\tau) \sim \tau^{-\alpha}
\label{Ptaualpha}
\end{equation}

\noindent with exponent $\alpha\approx1$, independent of the nature of the
activity. Given that for these activity patterns we collected data for
thousands of users, we need to calculate the distribution of the exponent
$\alpha$ determined separatelly for each user whose activity level exceeds
a certain threshold ({\it i.e.} avoiding users that have too few events to
allow a meaningful determination of $P(\tau)$). As Fig. \ref{F:fig2}e-g
shows, we find that the distribution of the exponents is peaked around
$\alpha=1$.

The scattering around $\alpha=1$ in the measured exponents could have two
different origins. First, it is possible that each user is characterized
by a different scaling exponent $\alpha$. Second, each user could have the
same exponent $\alpha =1$, but given the fact that the available dataset
captures only a finite time interval from one month to several months,
with at best a few thousand events in this interval, there are
uncertainties in our ability to determine numerically the exponent
$\alpha$. To demonstrate that such data incompleteness could indeed
explain the observed scattering, in Figs. \ref{F:fig2}h and \ref{F:fig2}l
we show the result of a numerical experiment, in which we generated 10,000
time series, corresponding to 10,000 independent users, the interevent
time of the events for each user being taken from the same distribution
$P(\tau) \sim \tau^{-1}$. The total length in time of each time series was
chosen to be $1,000,000$. We then used the automatic fitting algorithm
employed earlier to measure the exponents in Figs.  \ref{F:fig2}e-g to
determine numerically the exponent $\alpha$ for each user. In principle
for each user we should observe the same exponent $\alpha=1$, given that
the datasets were generated in an identical fashion. In practice, however,
due to the finite length of the data, each numerically determined exponent
is slightly different, resulting in the histogram shown in Fig.
\ref{F:fig2}h. As the figure shows, even in this well controlled situation
we observe a scattering in the measured exponents, obtaining a
distribution similar to the one seen in Figs. \ref{F:fig2}e-g. The longer
the time series, the sharper the distribution is (Fig. \ref{F:fig2}l),
given that the exponent $\alpha$ can be determined more accurately.

The distributions obtained for the three studied datasets are not as well
controlled as the one used in our simulation: while the length of the
observation period is the same for each user, the activity level of the
users differs widely. Indeed, as we show in Fig. \ref{F:fig2}i-k, the
activity distribution of the different users, representing the number of
events recorded for each user, also spans several orders of magnitude,
following a fat tailed distribution. Thus the degree of scattering of the
measured exponent $\alpha$ is expected to be more significant than seen in
Fig \ref{F:fig2}h and l, since we can determine the exponent accurately
only for very active users, for which we have a significant number of
datapoints. Therefore, the obtained results are consistent with the
hypothesis that each user is characterized by a scaling exponent in the
vicinity of $\alpha=1$, the difference in the numerically measured
exponent values being likely rooted in the finite number of events we
record for each user in the datasets. This conclusion will be eventually
corroborated by our modeling efforts, that indicate that the exponents
characterizing human behavior take up discrete values, one of which,
provide the empirically observed $\alpha=1$.

\begin{figure}
\centerline{\includegraphics[width=3.2in]{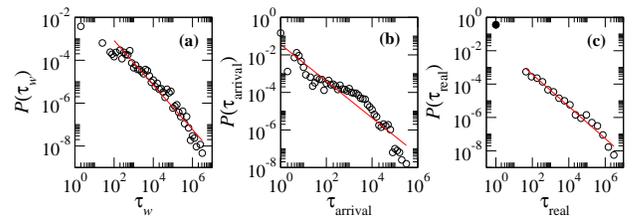}}

\caption{Distribution of the response and arrival time intervals of the
email user shown in Fig. \ref{F:fig2}b. (a) Given two email users A and B,
the response times of user A to B are the time intervals between A
receiving an email from B and A sending an email to B. The response time
distribution of user A is then computed taking into account the response
times to all users he/she communicates with. The continuous line is a
power law fit with exponent $\alpha=1.0$. (b) Given an user A, the
inter-arrival times are the time intervals between the two consecutive
arrivals of an email to user A, independently of the sender. The arrival
time distribution of user A is obtained taking into account all the
inter-arrival times for that user. The continuous line is a power law fit
with exponent 0.98. (c) The real waiting time distribution of an email in
a user's priority list, where $\tau_{\rm real}$ represents the time
between the time the user first sees an email and the time she sends a
reply to it. The black symbol shown in the upper left corner represents
the messages that were replied to right after the user has noticed it.}

\label{fig:arrival:response}
\end{figure}

As we will see in the following sections, an important measure of the
human activity patterns is the waiting time, $\tau_w$, representing the
amount of time a task waits on an individual's priority list before being
executed. For the email dataset, given that we know when a user receives
an email from another user and the time it sends the next email back to
her, we can determine the email's waiting or response time. Therefore, we
define the waiting time as the difference between the time user A receives
an email from user B, and the time A sends an email to user B. In looking
at this quantity we should be aware of the fact that not all emails A
sends to B are direct responses to emails received from B, thus there are
some false positives in the data that could be filtered out only by
reading the text of each email (which is not possible in the available
datasets). We have measured the empirically obtained waiting time
distribution in the email dataset, finding that the distribution of the
response times indeed follows a power law with exponent $\alpha=1$ (Fig.
\ref{fig:arrival:response}a).

\begin{figure}

\centerline{\includegraphics[width=3.2in]{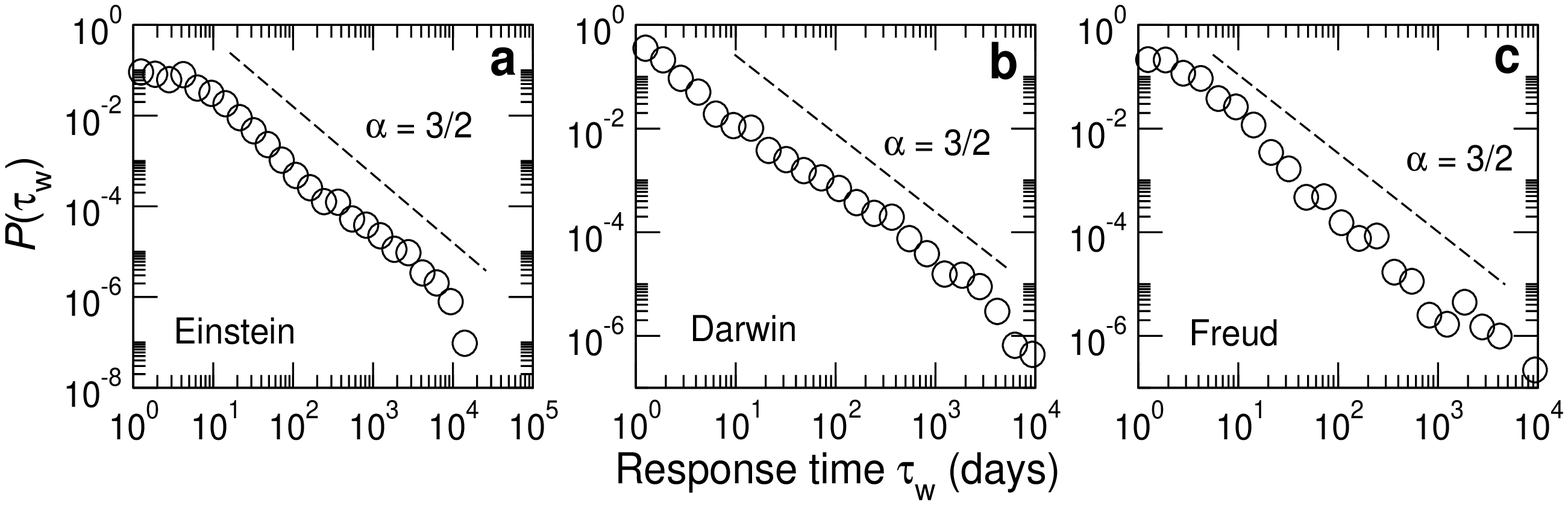}}

\caption{Distribution of the response times for the letters replied to by
Einstein, Darwin and Freud, as indicated on each plot. Note that the
distributions are well approximated with a power law tail with exponent
$\alpha=3/2$. While for Darwin and Einstein the datasets provide very good
statistics (the power law regime spanning 4 orders of magnitude), the plot
corresponding to Freud's responses is not so impressive, yet still being
well approximated by the power law distribution. Note that while in most
cases the identified reply is indeed a response to a received letter,
there are exceptions as well: many of the very delayed replies represent
the renewal of a long lost relationship.}

\label{fig:letters}
\end{figure}

\subsection{The $\alpha=3/2$ universality class: The correspondence of
Einstein, Darwin and Freud} \label{sec:letters}

In the case of the correspondence patterns of Einstein and Darwin we will
focus on the response time of the authors, partly because we will see
later that this has the most importance from the modeling perspective. As
shown in Fig. \ref{fig:letters}, the probability that a letter will be
replied to in $\tau_w$ days is well approximated by a power law
(\ref{Ptaualpha}) with $\alpha=3/2$, the scaling spanning four orders of
magnitude, from days to years.  Note that this exponent is significantly
different from $\alpha=1$ observed in the earlier datasets, and we will
show later that modeling efforts indeed establish $\alpha=3/2$ as a
scaling exponent characterizing human dynamics.

The dataset allows us to determine the interevent times as well,
representing the time interval between two consecutive letters sent by
Einstein, Darwin or Freud to any recipient. We find that the interevent
time distribution is also heavy tailed, albeit the quality of scaling is
not as impressive as we observe for the response time distribution. This
is due to the fact that we do not know the precise time when the letter is
written (in contrast with the email, which is known with second
resolution), but only the day on which it was mailed. Given that both
Einstein and Darwin wrote at least one letter most days, this means that
long interevent times are rarely observed. Furthermore, owing to the long
observational period (over 70 years), the overall activity pattern of the
two scientists has changed significantly, going from a few letters per
year to as many 400-800 letters/year during the later, more famous phase
of their professional life. Thus the interevent time, while it appears to
follow a power law distribution, it is by no means stationary. More
stationarity is observed, however, in the response time distribution.

\subsection{The stock broker activity pattern} 
\label{sec:broker}

For the stock broker we again focus on the interevent time distribution,
finding that the best fit follows $P(\tau) \sim \tau^{-\alpha}
\exp(-\tau/\tau_0)$ with $\alpha=1.3$ and $\tau_0=76$ min (see Fig. 1d).  
This value is between $\alpha=1$ observed for the users in the first three
other datasets and $\alpha=3/2$ observed for the correspondence patterns.
Yet, given the scattering of the measured exponents, it is difficult to
determine if this represents a standard statistical deviation from
$\alpha=1$ or $\alpha=3/2$, the two values expected by the modeling
efforts (see Sects. \ref{sec:cobham} and \ref{sec:barabasi}), or it stands
as evidence for a new universality class. At this point we believe that
the former case is valid, something that can be decided only once data for
more users will become available. The exponential cutoff is not
inconsistent with the modelling efforts either: as we will show in
Appendix \ref{general}, such cutoffs are expected to accompany all human
activity patterns with $\alpha<2$.

\subsection{Qualitative differences between heavy tailed and Poisson 
activity patterns}

The heavy tailed nature of the observed interevent time distribution
has clear visual signatures. Indeed, it implies that an individual's
activity pattern has a bursty character: short time intervals with
intensive activity (bursts) are separated by long periods of no
activity (Figs. 1d-f). Therefore, in contrast with the relatively
uniform activity pattern predicted by the Poisson process, for a
heavy tailed process very dense successions of events (bursts) are
separated by very long gaps, predicted by the slowly decaying tail
of the power law distribution. This bursty activity pattern agrees
with our experience of an individual's normal email usage pattern:
during a single session we typically send several emails in quick
succession, followed by long periods of no email activity, when we
focus on other activities.

\section{Capturing human dynamics: queuing models}
\label{sec:model}

The empirical evidence discussed in the previous section raises
several important questions: Why does the Poisson process fail to
capture the temporal features of human activity? What is the origin
of the observed heavy tailed activity patterns in human dynamics? To
address these questions we need to inspect closely the processes
that contribute to the timing of the events in which an individual
participates.

Most of the time humans face simultaneously several work, entertainment,
and family related responsibilities. Indeed, at any moment an individual
could choose to participate in one of several tasks, ranging from shopping
to sending emails, making phone calls, attending meetings or talks, going
to a theater, getting tickets for a sports event, and so on. To keep track
of the various responsibilities ahead of them, individuals maintain a {\it
to do} or {\it priority} list, recording the upcoming tasks. While this
list is occasionally written or electronically recorded, in many cases it
is simply kept in memory. A priority list is a dynamic entity, since tasks
are removed from it after they are executed and new tasks are added
continuously.  The tasks on the list compete with each other for the
individual's time and attention. Therefore, task management by humans is
best described as a queuing process \cite{queue-cohen,gross98}, where the
queue represents the tasks on the priority list, the server is the
individual which executes them and maintains the list, and some selection
protocol governs the order in which the tasks are executed. To define the
relevant queuing model we must clarify some key features of the underlying
queuing process, ranging from the arrival and service processes to the
nature of the task selection protocol, and the restrictions on the queue
length \cite{queue-cohen}. In the following we discuss each of these
ingredients separately, placing special emphasis on their relevance to
human dynamics.

{\it Server:} The server refers to the individual (or agent) that 
maintains the queue and executes the tasks. In queuing theory we can have 
one or several servers in parallel (like checkout counters in a 
supermarket). Human dynamics is a {\it single server} process, capturing 
the fact that an individual is solely responsible for executing the tasks 
on his/her priority list.

{\it Task Arrival Pattern:} The arrival process specifies the statistics
of the arrival of new tasks to the queue. In queuing theory it is often
assumed that the arrival is a Poisson process, meaning that new tasks
arrive at a constant rate $\lambda$ to the queue, randomly and
independently from each other. We will use this approximation for human
queues as well, assuming that tasks land at random times on the priority
list. If the arrival process is not captured by a Poisson distribution, it
can be modeled as a renewal process with a general distribution of
interarrival times \cite{queue-cohen}. For example, our measurements
indicate that the arrival time of emails follows a heavy tailed
distribution, thus a detailed modeling of email based queues must take
this into account. We must also keep in mind that the arrival rate of the
tasks to the list is filtered by the individual, who decides which tasks
to accept and place on the priority list and which to reject. In principle
the rejection of a task is also a decision process that can be modeled as
a high priority short lived task.

{\it Service process:} The service process specifies the time it takes for
a single task to be executed, such as the time necessary to write an
email, explore a web page or read a book. In queuing theory the service
process is often modeled as a Poisson process, which means that the
distribution of the time devoted to the individual tasks has the
exponential form (\ref{PPoisson}). However, in some applications the
service time may follow some general distribution. For example, the size
distribution of files transmitted by email is known to be fat tailed
\cite{Crovella,Mitzenmacher}, suggesting that the time necessary to review
(read) them could also follow a fat tailed distribution. In queuing theory
it is often assumed that the service time is independent of the task
arrival process or the number of tasks on the priority list. While we
adopt this assumption here as well, we must also keep in mind that the
service time can decrease if too many tasks are in the queue, as humans
may devote less time to individual tasks when they have many urgent things
to do.

{\it Selection protocol or queue discipline:} The selection protocol
specifies the manner in which the tasks in the queue are selected for
execution. Most human initiated events require an individual to weight and
prioritize different activities. For example, at the end of each activity
an individual needs to decide what to do next: send an email, do some
shopping or place a phone call, allocating time and resources for the
chosen activity. Normally individuals assign to each task a priority
parameter, which allows them to compare the urgency of the different tasks
on the list. The time a task waits before it is executed depends on the
method the agent uses to choose the task to be executed next. In this
respect three selection protocols are particularly relevant for human
dynamics:

({\it i}) The simplest is the first-in-first-out (FIFO) protocol,
executing the tasks in the order they were added to the list. This
is common in service oriented processes, like the
first-come-first-serve execution of orders in a restaurant or
getting help from directory assistance and consumer support.

({\it ii}) The second possibility is to execute the tasks in a
random order, irrespective of their priority or time spent on the
list. This is common, for example, in educational settings, when
students are called on randomly, and in some packet routing
protocols.

({\it iii})  In most human initiated activities task selection is not
random, but the individual tends to execute always the highest priority
item on his/her list. The resulting execution dynamics is quite different
from ({\it i}) and ({\it ii}): high priority tasks will be executed soon
after their addition to the list, while low priority items will have to
wait until all higher priority tasks are cleared, forcing them to stay
longer on the list. In the following we show that this selection
mechanism, practiced by humans on a daily basis, is the likely source of
the fat tails observed in human initiated processes.

{\it Queue Length or System Capacity:} In most queuing models the queue
has an infinite capacity and the queue length can change dynamically,
depending on the arrival and the execution rate of the individual tasks.  
In some queuing processes there is a physical limitation on the queue
length. For example, the buffers of Internet routers have finite capacity,
so that packets arriving while the buffer is full are systematically
dropped. In human activity one could argue that, given the possibility to
maintain the priority list in a written or electronic form, the length of
the list has no limitations. Yet, if confronted with too many
responsibilities, humans will start dropping some tasks and not accept
others. Furthermore, while keeping track of a long priority list is not a
problem for an electronic organizer, it is well established that the
immediate memory of humans has finite capacity of about seven tasks
\cite{Miller,baddeley94}. In other words, the number of priorities we can
easily remember, and therefore the length of our priority list, is
bounded. These considerations force us to inspect closely the difference
between finite and an unbounded priority lists, and the potential
consequences of the queue length on the the waiting time distribution.

In this paper we follow the hypothesis that the empirically observed heavy
tailed distributions originate in the queuing process of the tasks
maintained by humans, and seek appropriate models to explain and quantify
this phenomenon. Particularly valuable are queuing models that do not
contain power law distributions as inputs, and yet generate a heavy tailed
output. In the following we will focus on priority queues, reflecting the
fact that humans most likely choose the tasks based on their priority for
execution.

In the empirical datasets discussed in Sect \ref{sec:empirical} we focused
on both the interevent time and the waiting time distribution of the tasks
in which humans participate. In the following two sections we focus on the
{\it waiting time} of a task on the priority list rather than the
interevent times. In this context he waiting time, $\tau_w$, represents
the time difference between the arrival of a task to the priority list and
its execution, thus it is the sum of the time a task waits on the list and
the time devoted to executing it. In Sect. \ref{sec:wait_interevent} we
will return to the relationship between the empirically observed
interevent times and the waiting times predicted by the discussed models.

\section{Models with variable queue length: $\alpha=3/2$ universality 
class}

\label{sec:cobham}

Our first goal is to explore the behavior of priority queues in which
there are no restrictions on the queue length. Therefore, in these models
an individual's priority list could contain arbitrary number of tasks. As
we will show below, such models offer a good approximation to the surface
mail correspondence patterns, such as that observed in the case of
Einstein, Darwin and Freud (see Sect. \ref{sec:letters}). Therefore, we
will construct the models with direct reference to the the datasets
discussed in Sect.  \ref{sec:empirical}. We
assume that letters arrive at rate $\lambda$ following a Poisson process
with exponential arrival time distribution.  Replacing letters with tasks,
however, provides us a more general model, in principle applicable to any
human activity. The responses are written at rate $\mu$, reflecting the
overall time a person devotes to his correspondence. Each letter is
assigned a discrete priority parameter $x=1,2,\ldots,r$ upon arrival, such
that always the highest priority unanswered letter (task) will be always
chosen for a reply. The lowest priority task will have to wait the longest
before execution, and therefore it dominates the waiting time probability
density for large waiting times.  This model was introduced in 1964 by
Cobham \cite{Cobham} to describe some manufacturing processes. Most of the
analytical work in queuing theory has concentrated on the waiting time of
the lowest priority task, finding that the waiting time distribution
follows \cite{abate97}

\begin{equation}
P(\tau_w) \sim A\tau_w^{-3/2}\exp \left( -\frac{\tau_w}{\tau_0} \right)\ ,
\label{w0}
\end{equation}

\noindent where $A$ and $\tau_0$ are functions of the model
parameters, the characteristic
waiting time $\tau_0$ being given by

\begin{equation}
\tau_0 = \frac{1}{\mu \left( 1 - \sqrt{\rho} \right)^2 }\ ,
\label{tau0:cobham}
\end{equation}

\noindent where $\rho=\lambda/\mu$ is the traffic intensity. Therefore,
the waiting time distribution is characterized by a power law decay with
exponent $\alpha=3/2$, combined with an exponential cutoff.

\begin{figure}

\centerline{\includegraphics[width=3.2in]{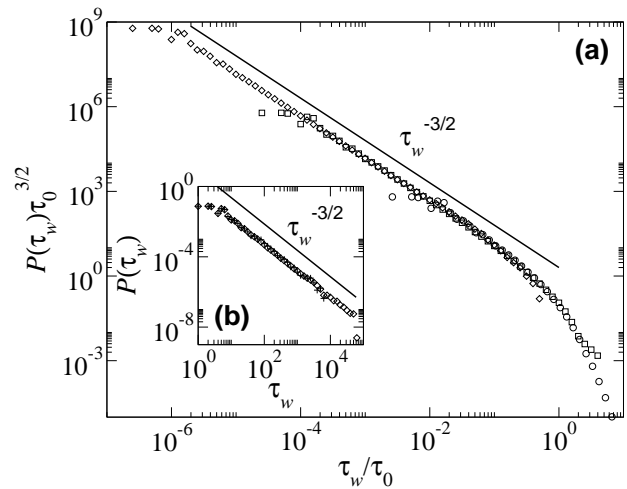}}

\caption{Waiting time distribution for tasks in the queueing model
discussed in Sect. \ref{sec:cobham} with continuous priorities. The
numerical simulations were performed as follows: At each step we generate
an arrival $\tau_a$ and service time $\tau_s$ from an exponential
distribution with rate $\lambda$ and $\mu$, respectively. If
$\tau_a<\tau_s$ or there are no tasks in the queue then we add a new task
to the queue, with a priority $x\in[0,1]$ from uniform distribution, and
update the time $t\rightarrow t+\tau_a$. Otherwise, we remove from the
queue the task with the largest priority and update the time $t\rightarrow
t+\tau_s$. The waiting time distribution is plotted for three
$\rho=\lambda/\mu$ values: $\rho=0.9$ (circles), $\rho=0.99$ (squares) and
$\rho=0.999$ (diamonds).  The data has been rescaled to emphasize the
scaling behavior $P(\tau_w)=\tau_w^{-3/2}f(\tau_w/\tau_0)$, where
$\tau_0\sim (1-\sqrt{\rho})^{-2}$. In the inset we plot the distribution
of waiting times for $\rho=1.1$, after collecting up to $10^4$ (plus) and
$10^5$ (diamonds) {\it executed} tasks, showing that the distribution of
waiting times has a power law tail even for $\rho>1$ (supercritical
regime). Note, however, that in this regime a high fraction of tasks are
never executed, staying forever on the priority list whose length
increases linearly with time, a fact that is manifested by a shift to the
right of the cutoff of the waiting time distribution.}

\label{cobham-hpf}
\end{figure}

The model can be extended to the case where the priorities are not
discrete, but take up continuous values $0\leq x<\infty$ from an arbitrary
$\eta(x)$ distribution. The Laplace transform of the waiting time
distribution for this case has been calculated in Ref. \cite{queue-cohen},
but the resulting equation is difficult to invert, forcing us to study the
model numerically (Fig. \ref{cobham-hpf}). The natural control parameter
is $\rho=\lambda/\mu$, allowing us to distinguish three qualitatively
different regimes:

{\it Subcritical regime}, $\rho<1$: Given that the arrival rate of the
tasks is smaller than the execution rate, the queue will be often empty.
This significantly limits the waiting time, most tasks being executed soon
after their arrival. The simulations indicate that the waiting time
distribution exhibits an asymptotic scaling behavior consistent with
(\ref{w0}) (Fig. \ref{cobham-hpf}). While in the $\rho\rightarrow0$ limit
we observe mainly the exponential decay, as $\rho$ approaches 1 a power
law regime with exponent $\alpha=3/2$ emerges, combined with the
exponential cutoff.

{\it Critical regime}, $\rho=1$ : When the arrival and the response rate
of the letters are equal, according to (\ref{w0}) and (\ref{tau0:cobham})  
we should observe a power law waiting time distribution with $\alpha=3/2$
(Fig. \ref{cobham-hpf}). This regime would imply that, for example, Darwin
responds to all letters he receive, which is not the case,
given that their response rate is 0.32 (Darwin), 0.24 (Einstein) and 0.31
(Freud)  \cite{joao05}. In this case it is easy to show that the queue
length performs a one-dimensional random walk bounded at $l=0$. These
fluctuations in the queue length will limit the waiting time distribution,
as the tasks will wait at most as long as it takes for the queue length to
return to $l=0$. Therefore, the waiting time distribution will have as
upper bound the return time distribution of a one-dimensional random walk.
It is known, however, that the return time distribution of a random walker
follows $P(t)\sim t^{-3/2}$ \cite{redner01,havlinRW}, which is the origin
of the $3/2$ exponent in Eq. (\ref{Ptaualpha}). This argument indicates
that (\ref{w0}) is related to the fluctuations in the length of the
priority list.

{\it Supercritical regime}, $\rho>1$ : Given that in this regime the
arrival rate exceeds the response rate, the average queue length grows
linearly as $\langle l(t)\rangle=(\lambda-\mu)t$. Therefore, a $1-1/\rho$
fraction of the letters is never responded to, waiting indefinitely in the
queue. Given Darwin, Einstein and Freud's small response rate, this regime
captures best their correspondence pattern. We can measure the waiting
time for each letter that is responded to. In Fig. \ref{cobham-hpf} we
show the waiting time probability density obtained from numerical
simulations, indicating that it follows a power law with exponent
$\alpha=3/2$. Thus the supercritical regime follows the same scaling
behavior as the critical regime, but only for the letters that are
responded to. The rest of the letters wait indefinitely in the list
($\tau_w=\infty$).

While the discussed model can indeed generate power law waiting time
distributions, a critical comparison with the empirical datasets reveals
some notable deficiencies. First, a power law distribution emerges only in
the critical ($\rho=1$) and the supercritical ($\rho>1$) regimes. The
critical regime requires a careful tuning of the human execution rate, so
that the execution and the arrival rates are exactly the same. In
contrast, for $\rho >1$ no tuning is necessary, but the number of tasks on
the list increases linearly with time, thus many tasks are never executed.  
This limit is probably the most realistic for human dynamics: we often
take on tasks that we never execute, and technically stay on our priority
list forever. As we discussed above, this is documentedly the case for
Einstein, Darwin and Freud, who answer only a fraction of their letters.  
However, we must not overlook the second important feature of the
discussed model: the only exponent it can predict is $\alpha=3/2$, rooted
in the fluctuations of the queue length. While this fully agrees with the
correspondence patterns of Einstein, Darwin and Freud, it is significantly
higher than the values observed in the empirical data discussed in Sect.
\ref{sec:alpha1} on web browsing, email communications or library visits,
which we found to be scattered around $\alpha=1$.

\section{Models with fixed queue length: $\alpha=1$ universality class} 
\label{sec:barabasi}

To understand the limitations of the model discussed in the previous
section we must remember that when the arrival and execution rates are
equal ($\rho=1$) the length of the priority list follows a random walk,
and can thus occasionally take up very large values. The situation is even
worse for $\rho>1$, when the queue length increases linearly with time.
Therefore, according to the model an individual must have the capacity to
keep track of hundreds or thousands of tasks at the same time. This may be
appropriate for surface mail, where the letters pile on our desk until
replied to. In contrast, there is extensive evidence from the psychology
literature that the number of tasks humans can easily keep in their short
term memory is bounded \cite{Miller}, therefore it is unrealistic that we
will remember hundreds or thousands of tasks at any given time. This
forces us to inspect a model in which the length of the priority list
remains unchanged \cite{barabasi05}, a new task being added only when an
old task is removed from the list (executed).

We assume that an individual mantains a priority list with $L$ tasks, each
task being assigned a priority parameter $x_i, ~i=1,..., L$, chosen from
an $\eta(x)$ distribution. At each time step with probability $p$ the
individual selects the highest priority task and executes it, removing it
from the list. At that moment a new task is added to the list, its
priority $x_i$ is again chosen from $\eta(x)$, thus the length $L$ of the
list remains unchanged. With probability $1-p$ the individual executes a
randomly selected task, independent of its priority.  The $p \to 1$ limit
of the model describes the deterministic highest-priority-first protocol,
when always the highest priority task is chosen for execution, while $p
\to 0$ corresponds to the random choice protocol, introduced to mimic the
fact that humans occasionally select some low priority items for
execution, before all higher priority items are executed. In the model
time is discrete, each task execution corresponding to one unit of time.
Implicit in this assumption is the approximation that the service time
distribution follows a delta function, {\it i.e.}, each task takes one
unit time to execute.

\begin{figure}
\centerline{\includegraphics[width=3.2in]{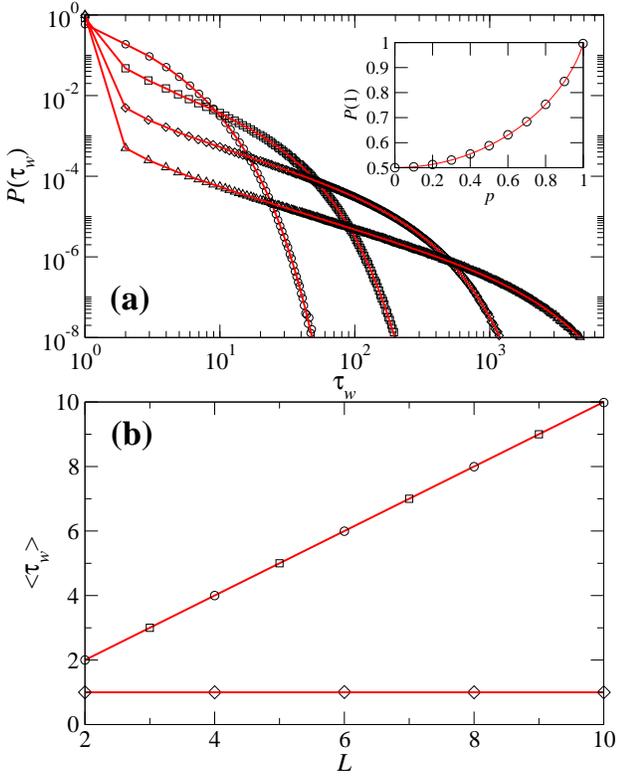}}

\caption{(a) Waiting time probability distribution function for the model
discussed in Sect. \ref{sec:barabasi} for $L=2$ and a uniform new task
priority distribution function, $\eta(x)=1$, in $0\leq x\leq 1$, as
obtained from (\ref{ptau}) (lines) and numerical simulations (symbols),
for $p=0.9$ (squares), $p=0.99$ (diamonds) and $p=0.999$ (triangles). The
inset shows the fraction of tasks with waiting time $\tau=1$, as obtained
from (\ref{ptau}) (lines) and numerical simulations (symbols). (b) Average
waiting time of executed tasks vs the list size as obtained from
(\ref{tauaveL}) (lines) and numerical simulations (symbols), for $p=0.0$
(squares), $p=0.999$ (circles) and $p=1$ (diamonds).}

\label{fig:l2}
\end{figure}

To understand the dynamics of the model we first study it via numerical
simulations with priorities chosen from a uniform distribution $x_i \in
[0,1]$. The simulations show that in the $p \to 1$ limit the probability
that a task spends $\tau_w$ time on the list has a power law tail with
exponent $\alpha=1$ (Fig. \ref{fig:l2}a). In the $p\to 0$ limit
$P(\tau_w)$ follows an exponential distribution (Fig. \ref{fig:l2}a), as
expected for the random selection protocol. As the typical length of the
priority list differs from individual to individual, it is important for
the tail of $P(\tau_w)$ to be independent of $L$. Numerical simulations
indicate that this is indeed the case: changes in $L$ do not affect the
scaling of $P(\tau_w)$ \cite{barabasi05}. The fact that the scaling holds
for $L=2$ as well indicates that it is not necessary to have a long
priority list: even if an individual balances only two tasks at the same
time, a bursty heavy tailed interevent dynamics will emerge. Next we focus
on the $L=2$ case, for which the model can be solved exactly, providing
important insights into its scaling behavior that can be generalized for
arbitrary $L$ values as well.

\subsection{Exact solution for $L=2$}
\label{sec:Barabasi:L2}

For $L=2$ the waiting time distribution can be determined exactly
\cite{vazquez05} (see Appendix \ref{app:l2}), obtaining

\begin{equation}
P(\tau_w) = \left\{
\begin{array}{ll}
1 - \frac{1-p^2}{4p} \ln \frac{1+p}{1-p}\ , & \tau_w=1\\
\\
\frac{1-p^2}{4p(\tau_w-1)} \left[ \left(\frac{1+p}{2}\right)^{\tau_w-1}
- \left(\frac{1-p}{2}\right)^{\tau_w-1} \right]\ , & 
\tau_w>1
\end{array}
\right.
\label{ptau}
\end{equation}

\noindent independent of $\eta(x)$ from which the task priorities
are selected. In the limit $p\rightarrow0$ from (\ref{ptau}) follows
that

\begin{equation}
\lim_{p\rightarrow0}P(\tau_w) = \left(\frac{1}{2}\right)^{-\tau_w}\ ,
\label{ptaup0}
\end{equation}

\noindent {\it i.e.} $P(\tau_w)$ decays exponentially, in agreement
with the numerical results (Fig. \ref{fig:l2}a). This limit
corresponds to the random selection protocol, where a task is
selected with probability $1/2$ in each step.  In the
$p\rightarrow1$ limit we obtain

\begin{equation}
\lim_{p\rightarrow1}P(\tau_w) = \left\{
\begin{array}{ll}
1 + {\cal O}\left(\frac{1-p}{2}\ln(1-p)\right)\ , & \tau_w=1\\
\\
{\cal O}\left(\frac{1-p}{2}\right) \frac{1}{\tau_w-1}\ , & \tau_w>1\ .
\end{array}
\right.
\label{ptaup1}
\end{equation}

\noindent In this case almost all tasks have a waiting time
$\tau_w=1$, being executed as soon as they were added to the
priority list. The waiting time of tasks that are not selected in
the first step follows a power law distribution, decaying with
$\alpha=1$. This behavior is illustrated in Fig. \ref{fig:l2}a by a
direct plot of $P(\tau_w)$ in (\ref{ptau}) for a uniform
distribution $\eta(x)$ in $0\leq x\leq1$. For $p<1$ the $P(\tau_w)$
distribution has an exponential cutoff, which can be derived from
(\ref{ptau}) after taking the $\tau_w\rightarrow\infty$ limit with
$p$ fixed, resulting in

\begin{equation}
P(\tau_w) \sim \frac{1-p^2}{4}\frac{1}{\tau_w}
\exp\left( - \frac{\tau_w}{\tau_0}  \right)\ ,
\label{ptau00}
\end{equation}

\noindent where

\begin{equation}
\tau_0 = \left( \ln \frac{2}{1+p} \right)^{-1}\ .
\label{tau0}
\end{equation}

\noindent When $p\rightarrow1$ we obtain that $\tau_0\rightarrow\infty$ 
and, therefore, the exponential cutoff is shifted to higher $\tau_w$ 
values, while the power law behavior $P(\tau_w)\sim 1/\tau_w$ becomes more 
prominent. The $P(\tau_w)$ curve systematically shifts, however, to lower 
values for $\tau_w>1$, indicating that the power law applies to a 
vanishing task fraction (see Fig. \ref{fig:l2}a and (\ref{ptau00})). In 
turn, $P(1)\rightarrow1$ when $p\rightarrow1$, corroborated by the direct 
plot of $P(1)$ as a function of $p$ (see inset of Fig. \ref{fig:l2}a).

\subsection{Numerical results for $L>2$}
\label{sec:Barabasi:Lg2}

Based on the results discussed above, the overall behavior of the model
with a uniform priority distribution can be summarized as follows. For
$p=1$, corresponding to the case when {\it always} the highest priority
task is removed, the model does not have a stationary state. Indeed, each
time the highest priority task is executed, there is a task with smaller
priority $x_m$ left on the list.  With probability $1-x_m$ the newly added
task will have a priority $x^\prime_m$ larger than $x_m$, and will be
executed immediately. With probability $x_m$, however, the new task will
have a smaller priority, in which case the older task will be executed,
and the new task will become the `resident' one, with a smaller priority
$x_m'<x_m$. For a long period all new tasks will be executed right away,
until an another task arrives with probability $x_m^{\prime\prime}$ that
again pushes the non-executed priority to a smaller value
$x_m^{\prime\prime}<x_m^{\prime}$. Thus with time the priority of the
lowest priority task will converge to zero, $x_m(t)\rightarrow0$, and thus
with a probability converging to one the new task will be immediately
executed. This convergence of $x_m$ to zero implies that for $p=1$ the
model does not have a stationary state. A stationary state develops,
however, for any $p<1$, as in this case there is always a finite chance
that the lowest priority tasks will also be executed, thus the value of
$x_m$ will be reset, and will converge to some $x_m(p)>0$. This
qualitative description applies for arbitrary $L>2$ values.

\begin{figure}
\centerline{\includegraphics[width=3.2in]{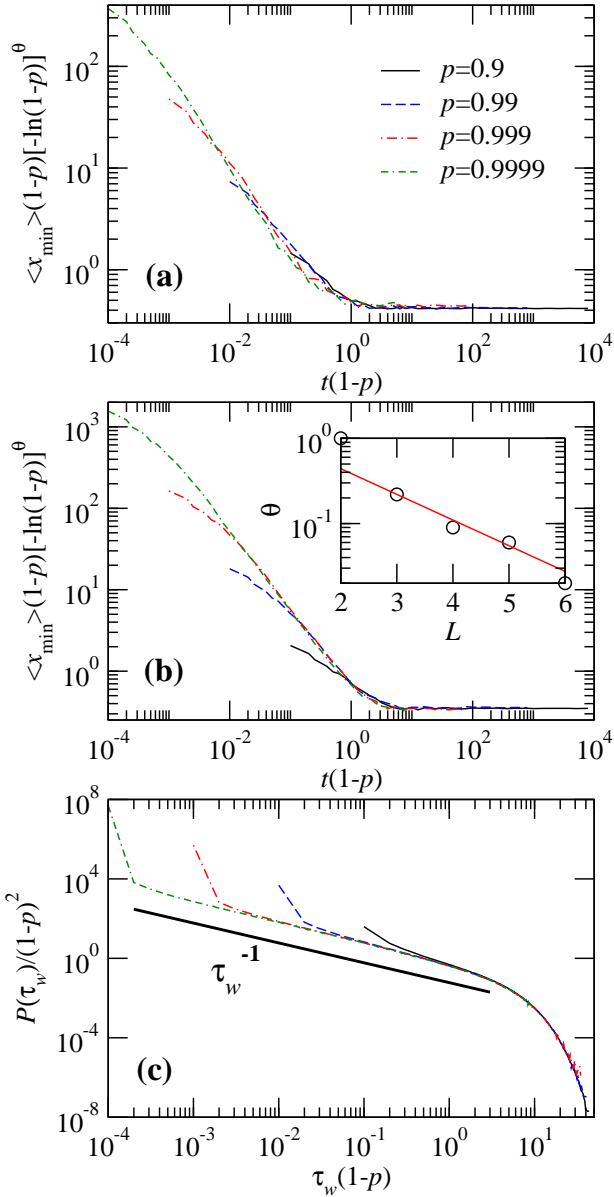}}

\caption{Rescaled plot of the average priority of the lowest task priority
in the list for $L=2$ (a) and $L=3$ (b) and different values of $p$ (see
legend). The inset in (b) shows the exponent $\theta_L$ for different $L$
(points), indicating that $\theta_L=\theta_3/2^{L-3}$ for $L>2$
(continuous line). (c) Rescaled plot of the waiting time distribution for
$L=3$.  Similar plots are obtained for larger vales of $L$ (data not
shown).}

\label{fig:scaling}
\end{figure}

To quantify this qualitative picture we studied numerically the $L>2$ case 
assuming that $\eta(x)$ is uniformly distributed in the $0\leq x\leq1$ 
interval. To investigate how fast the system approaches the stationary 
state we compute the average priority of the lowest priority task in the 
queue, $\langle x_{\rm min}(t)\rangle$ (see Fig. \ref{fig:scaling}a,b) 
since it represents a lower bound for the average of any other priorities 
on the list. We find that for any $L$ values $\langle x_{\rm 
min}(t)\rangle$ decreases exponentially up to a time scale $t_0$, when it 
reaches a stationary value $\langle x_{\rm min}(\infty)\rangle$. The 
numerical simulations indicate that

\begin{equation}
t_0 \sim \frac{1}{1-p}\ ,
\label{t0}
\end{equation}

\begin{equation}
\langle x_{\rm min}(\infty)\rangle \sim
(1-p)[-\ln(1-p)]^{\theta_L}\ .
\label{xmin}
\end{equation}

\noindent For $L=2$ can calculate $\langle x_{\rm min}(\infty)\rangle$
exactly \cite{vazquez05}, obtaining

\begin{eqnarray}
\langle x_{\rm min}(\infty)\rangle &=& \frac{1-p}{2p} \left(
\frac{1+p}{2p}\ln\frac{1+p}{1-p} -1 \right)\nonumber\\
& \approx & \frac{1-p}{2}[-\ln(1-p)]\ ,
\label{xminl2}
\end{eqnarray}

\noindent and therefore $\theta_2=1$. For $L>2$ we determined $\theta_L$
from the best data collapse, obtaining the values shown in the inset of
Fig. \ref{fig:scaling}b, indicating that $$
\theta_L=\frac{\theta_3}{2^{L-3}} \ , $$ where $\theta_3=0.22$ is the
value of $\theta_L$ for $L=3$. These results support our qualitative
discussion, indicating that for all $L\geq2$ and $0\leq p<1$ values the
system reaches a stationary state.

Finally we measured the waiting time distribution after the system has
reached the stationary state. The results for $L=3$ are shown in Fig.
\ref{fig:scaling}c, and similar results were obtained for other $L>2$
values. The data collapse of the numerically obtained $P(\tau)$ indicates
that

\begin{equation}
P(\tau) \sim (1-p)^2\frac{1}{\tau}
\exp\left( - \frac{\tau}{\tau_0} \right)\ ,
\label{ptaulg2}
\end{equation}

\noindent when $L>2$ and $\tau\gg1$, where

\begin{equation}
\tau_0 \sim \frac{1}{1-p}
\label{tau0lg2}
\end{equation}

\noindent in the $p\rightarrow1$ limit. The simulations indicate that the
model's behavior for $L>2$ is qualitatively similar to the behavior
derived exactly for $L=2$, but different scaling parameters characterize
the scaling functions. For any $L\geq2$, however, the waiting times scale
as $P(\tau_w) \sim \tau_w^{-1}$, {\it i.e.} we have $\alpha=1$.

\subsection{Comparison with the empirical data}

As the results in the previous subsections show, the model proposed to
account for the $\alpha=1$ universality class has some apparent problems.
Indeed, for truly deterministic execution ($p=1$) the model does not have
a stationary state. The problem was cured by introducing a random task
execution ($p<1$), which leads to stationarity. In this case, however, a
$p$ dependent fraction of tasks are executed immediately, and only the
rest of the long lived tasks follow a power law. As $p$ converges to zero,
the fraction of tasks executed immediately diverges, developing a
significant gap between the power law regime, and the tasks displaying
$\tau=1$ waiting time. Is this behavior realistic, or represents an
artifact of the model? A first comparison with the empirical data would
suggest that this is indeed an artifact, as measurements shown in Fig.  
\ref{F:fig2} and \ref{fig:arrival:response} do not provide evidence of a
large number of tasks that are immediately executed. However, when
inspecting the measurement results we should keep in mind that they
represents the intervent times, and not the waiting times. In the case
when the waiting time can be directly measured, like in the email or mail
based correspondence, there is some ambiguity to the real waiting time.
Indeed, in the email data, for example, we have measured as waiting time
the time difference between the arrival of an email, and the response sent
to it.  While this offers an excellent approximation, from an individual's
or a priority queue's perspective this is not the real waiting time.
Indeed, consider the situation when an email arrives at 9:00 am, and the
recipient does not check her email until 11:56am, at which point she
replies to the email immediately. From the perspective of her priority
list the waiting time was less than a minute, as she replied as soon as
she saw the email. In our dataset, however, the waiting time will be 3
hours and 56 minutes. Thus the way we measured the waiting times cannot
identify the true waiting time of a task on a user's priority list. The
email dataset allows us, however, to get a much better approximation of
the real waiting times than we did before. Indeed, for an email $e_1$
received by user A we record the time $t_1$ it arrives, and then the time
$t_2$ of the first email sent by user A to any other user {\it after} the
arrival of the selected email. It will be this time from which we start
measuring the waiting time for email $e_1$. Thus if user A replies to
$e_1$ at time $t_3$, we consider that the email's waiting time
$\tau_{\rm real}=t_3-t_2$, instead of $t_3 -t_1$ considered in Fig
\ref{fig:arrival:response}a. The results, shown in Fig
\ref{fig:arrival:response}c, displays the same power law scaling with
$\alpha=1$ as we have seen in Fig \ref{fig:arrival:response}a, but in
addition there is a prominent peak at $\tau_{\rm real}=1$, cooresponding
to emails responded to immediately. Note that the peak's magnitude is
orders of magnitude larger than the probabilities displayed by the large
waiting times. This result suggests that what we could have easily
considered a model artifact in fact captures a common feature of email
communications.  Indeed, a high fraction of our emails is responded
immediately, right after our first chance to read them, as predicted by
the priority model discussed in this section. Are there models that can
provide the $\alpha=1$ universality class without the high fraction of
items executed imediatelly? While we have failed to come up with any
examples, we belive that developing such models could be quite valuable.

\section{Relationship between waiting times and interevent times}
\label{sec:wait_interevent}

As we discussed above, the empirical measurements provide either the
interevent time distribution $P(\tau)$ (Sects. \ref{sec:alpha1} and
\ref{sec:broker}) or the waiting time distribution $P(\tau_w)$ (Sect.  
\ref{sec:letters}) of the measured human activity patterns. In contrast
the model predicts only the waiting time $\tau_w$ of a task on an
individual's priority list. What is the relationship between the observed
interevent times and the predicted waiting times?  The basic thesis of our
paper is that the waiting times the various tasks experience on an
individual's priority list is responsible for the heavy tailed
distributions seen in the interevent times as well. The purpose of this
section is to discuss the relationship between the two quantities.

The model predictions, that the waiting time distribution of the tasks
follows a power law, is directly supported by one dataset in each
universality class: the email data and the correspondence data. As
discussed in Sect.  \ref{sec:empirical}, we have measured the waiting time
distribution for both datasets, finding that the distribution of the
response times indeed follows a power law with exponent $\alpha=1$ (email)
and $\alpha=3/2$ (correspondence mail)  as predicted by the models.
Therefore, the direct measurement of the waiting times are likely rooted
in the fat tailed response time distribution. For the other three
datasets, however, such as web browsing, library visits and stock
purchases, we cannot determine the waiting time of the individual events,
as we do not know when a given task is added to the individual's priority
list.

To explore the broader relationship between the waiting times and the
interevent times we must remind ourselves that while during the
measurements we are focusing on a specific task (like email), the models
assume the knowledge of {\it all} tasks that an individual is involved in.
Thus the empirical measurements offer only a selected subset of an
individual's activity pattern. To see the relationship between $\tau$ and
$\tau_w$ next we discuss two different approaches.

{\it Queueing of different task categories:} The first approach
acknowledges the fact that tasks are grouped in different categories of
priorities: we often do not keep in mind specific emails to be answered,
but rather remember that we need to check our email and answer whatever
needs attention. Similarly, we may remember a few things that we need to
shop for, but our priority list would often contain only one item: go to
the supermarket. When we monitor different human activity patterns, we see
the repetitive execution of these categories, like going to the library,
or doing emails, or browsing the web. Given this, one possible
modification of the discussed models would assume that the tasks we
monitor correspond to specific activity categories, and when we are done
with one of them, we do not remove it from the list, but we just add it
back with some changed priority. That is, checking our email does not mean
that we deleted email activity from our priority list, but only that next
has some different priority. If we monitor only one kind of activity, then
a proper model would be the following: we have $L$ tasks, each assigned a
given priority. After a task is executed, it will be reinserted in the
queue with a new priority chosen from the same distribution $\eta(x)$. If
we now monitor the time at which the different tasks exit the list, we
will find that the interevent times for the {\it monitored} tasks
correspond exactly to the waiting time of that task on the list. Note that
this conceptual model would work even if the tasks are not immediately
reinserted, but after some delay $\tau_d$. Indeed in this case the
interevent time will be $\tau=\tau_w+\tau_d$, and as long as the
distribution from which $\tau_d$ is selected from is bounded, the tail of
the interevent time distribution will be dominated by the waiting time
statistics.

{\it Interaction between individuals:} The timing of specific emails also
depends on the interaction between the individuals that are involved in an
email based communication.  Indeed, if user A gets an email from user B,
she will put the email into her priority list, and answer when she gets to
it. Thus the timing of the response depends on two parameters: the receipt
time of the email, and the waiting time on the priority list. Consider two
email users, A and B, that are involved in an email based conversation. We
assume that A sends an email to B as a response to an email B sent to A,
and viceversa. Thus, the interevent time between two consecutive emails
sent by user A to user B is given by $\tau=\tau_w^A+\tau_w^B$, where
$\tau_w^A$ is the waiting time the email experienced on user A's queue,
and $\tau_w^B$ is the waiting time of the response of user B to A's email.
If both users prioritize their tasks, then they both display the same
waiting time distribution, {\it i.e.} $P(\tau_w^A)\sim
(\tau_w^A)^{-\alpha}$ and $P(\tau_w^B)\sim (\tau_w^B)^{-\alpha}$. In this
case the interevent time distribution $P(\tau)$, which is observed
empirically if we study only the activity pattern of user A, follows also
$P(\tau)\sim\tau^{-\alpha}$. Thus the fact that users communicate with
each other turns the waiting time into an observable interevent times.

In summary, the discussed mechanisms indicate that the waiting time
distribution of the tasks could in fact drive the interevent time
distribution, and that the waiting time and the interevent time
distributions should decay with the same scaling exponent. In reality, of
course, the interplay between the two quantities can be more complex than
discussed here, and perhaps even better mapping between the two measures
could be found for selected activities. But these two mechanisms indicate
that if the waiting time distribution is heavy tailed, we would expect
that the interevent time distribution would be also affected by it.

\section{Discussion}
\label{sec:discussion}

{\it Universality classes:} As summarized in the introduction, the main
goal of the present paper was to discuss the potential origin of the heavy
tailed distributed interevent times observed in human dynamics. To start
we provided evidence that in five distinct processes, each capturing a
different human activity, the interevent time distribution for individual
users follow a power law. Our fundamental hypothesis is that the observed
interevent time distributions are rooted in the mechanisms that humans use
to decide when to execute the tasks on their priority list. To support
this hypothesis we studied a family of queuing models, that assume that
each task to be executed by an individual waits some time on the
individual's priority list and we showed that queuing can indeed generate
power law waiting time distributions. We find that a model that allows the
queue length to fluctuate leads to $\alpha=3/2$, while a model for which
the queue length is fixed displays $\alpha=1$. These results indicate that
human dynamics is described by at least two universality classes,
characterized by empirically distinguishable exponents. Note that while we
have classifed the models based on the limitations on the queue lenght, we
cannot exclude the existence of models with fixed queue lenght that scale
with $\alpha=3/2$, or models with fluctuating lenght that display scaling
with $\alpha=1$, or some other exponents.

In comparing these results with the empirical data, we find that email and
phone communication, web surfing and library visitation belong to
the $\alpha=1$ universality class. The correspondence patterns of
Einstein, Darwin and Freud offer convincing evidence for the relevance of
the $\alpha=3/2$ exponent, and the related universality class, for human
dynamics. In contrast the fourth process, capturing a stock broker's
activity, shows $\alpha=1.3$. Given, however, that we have data only for a
single user, this value is in principle consistent with the scattering of
the exponents from user to user, thus we cannot take it as evidence for a
new universality class.  One issue still remains without a satisfactory
answer: why does email and surface mail (Einstein, Darwin and Freud
datasets) belong to different universality classes? We can comprehend why
should the mail correspondence belong to the $3/2$ class: letters likely
pile on the correspondent's desk until they are answered, the
desk serving as an external memory, thus we do not require to remember 
them
all. But the same argument could be used to explain the scaling of email
communications as well, given that unanswered emails will stay in our
mailbox until we delete them (which is one kind of task execution).
Therefore one could argue that email based communication should also
belong to the $3/2$ universality class, in contrast with the empirical
evidence, that clearly shows $\alpha=1$ \cite{barabasi05,eckmann}.

Some difficulty in comparing the empirical data with the model predictions
is rooted in the fact that the models predict the waiting times, while for
many real systems only the interevent times can be measured. It is
encouraging, however, that for the email and the surface mail based
commnunication we were able to determine directly the waiting times as
well, and the exponents agreed with those determined from the interevent
times. In addition we argued that in a series of processes the waiting
time distribution determines the interevent time distribution as well (see
Sect. \ref{sec:wait_interevent}). This argument closes the loop of the
paper's logic, establishing the relevance of the discussed queueing models
to the datasets for which only interevent times could be measured. We do
not feel, however, that this argument is complete, and probably future
work will strengthen this link. In this respect two directions are
particularly promising. First, designing queueing models that can directly
predict the observed interevent times as well would be a major advance.
Second, establishing a more general link between the waiting time and
interevent times could also be of significant value.

The results discussed in this paper leave a number of issues unresolved.
In the following we will discuss some of these, outlining how answering
them could further our understanding of the statistical mechanics of human
driven processes.

{\it Tuning the universality class:} As we discussed above, the discussed
models provide evidence for two distinct universality classes in human
dynamics, with distinguishable exponents. The question is, are there other
universality classes, characterized by exponents different from 1 and 3/2?  
If other universality classes do exist, it would be valuable not only to
find empirical support for them, but also to identify classes of models
that are capable of predicting the new exponents.

In searching for new exponents we need to explore several different
directions. First, if one inserts some power law process into the queuing
model, that could tune the obtained waiting time distribution, and the
scaling exponents. There are different ways to achieve this. One method,
discussed in Appendix \ref{sec:preferential}, is based on the hypothesis
that while we always attempt to select the highest priority task,
circumstances or resource availability may not allow us to achieve this.  
For example, our highest priority may be to get cash from the bank, but we
cannot execute this task when the bank is closed, moving on to some lower
priority task. One way to account for this is to use a probabilistic
selection protocol, assuming that the probability to choose a task with
priority $x$ for execution in a unit time is $\Pi(x) \sim x^\gamma$, where
$\gamma$ is a parameter that interpolates between the random choice limit
(ii) ($\gamma=0$, $p=0$) and the deterministic case, when always the
highest priority item is chosen for execution (iii) ($\gamma=\infty$,
$p=1$). As shown in the Appendix \ref{sec:preferential}, the exponent
$\alpha$ will depend on $\gamma$ as

\begin{equation}
\alpha=1+\frac{1}{\gamma}\ .
\label{alphagamma}
\end{equation}

\noindent At this moment we do not have evidence that such preferential
selection process acts in human dynamics. However, detailed datasets and
proper measurement tools might help up decide this by measuring the
function $\Pi(x)$ directly, capturing the selection protocol. Such
measurements were possible for complex networks, where a similar function
drives the preferential attachment process
\cite{jeong03,barabasi1,review1,barabasi3,internet1,internet2}.

As we discussed above, the main goal of this paper was to demonstrate that
the queuing of the tasks on an individual's priority list can explain the
heavy tailed distributions observed in human activity patterns. To achieve
this, we focused on models with Poisson inputs, meaning that both the
arrival time and the execution time are bounded. In some situations,
however, the input distributions can be themselves heavy tailed. This
could have two origins: ({\it i}) Heavy tailed arrival time distribution:
As we show in Fig.  \ref{fig:arrival:response}b, there is direct evidence
for this in the email communication datasets: we find that the interevent
time of arriving emails can be roughly approximated with a power law with
exponent $\alpha_{in}=1$. ({\it ii}) The execution time could also be
heavy tailed, describing the situation when most tasks are executed very
rapidly, while a few tasks require a very long execution time. Evidence
for this again comes from the email system: the file sizes transmitted by
email are known to follow a heavy tailed distribution
\cite{Crovella,Mitzenmacher}.  Therefore, if we read every line of an
email, in principle the execution time should also be heavy tailed ({\it
i.e.} the time we actually take to work on the response, including reading
the original email). Note, however, that measurements failed to establish
a correlation between email size and the response time \cite{barabasi05}.
It is not particularly surprising that both ({\it i}) and ({\it ii}) would
significantly impact the waiting time distribution, generating a heavy
tailed distribution for the waiting times even when the behavior of the
model otherwise would be exponential, or change the exponent $\alpha$,
thus altering the model's universality class. Some aspects of this problem
were addressed recently by Blanchard and Hongler \cite{blanchard}.  
However, to illustrate the impact of the heavy tailed inputs in Appendix
\ref{app:service} we study the model of Sect. \ref{sec:cobham} with a
heavy tailed service time distribution $h(s)\sim s^{-\beta}$ with
$0<\beta<1$.

Finally, could the power law distributed arrival and execution times serve
as the proper explanation for the observed heavy tailed interevent time
distribution in human dynamics? Note that in a number of systems we
observe heavy tailed distributed events without evidence for power law
inputs. For example, the timing of the library visits or stock purchases
by brokers does not appear to be driven by any known power law inputs, and
they have negligible execution time compared with the average observed
interevent times.  Similarly, the beginning of online games or instant
messages is not driven by file sizes either, but only by the time
availability for playing a game or sending a message, which is mostly a
priority driven issue. Therefore, while it is important to understand the
impact of power law inputs on the scaling properties of various models,
attempts to explain the waiting times solely based on the heavy tailed
inputs only delegate the problem to an earlier cause (the origin of the
power law input).

{\it Potential model extensions:} Guided by the desire of constructing the
simplest models that capture the essence of task execution, we have
neglected many processes that are obviously present in human dynamics. For
example, we assumed that the priority of the tasks is assigned at the
moment the task was added to an individual's priority list, and remains
unchanged for the rest of the queuing process. In reality the priorities
themselves can change in time.  For example, many tasks have deadlines
\cite{blanchard}, and one could assume that a task's priority diverges as
the deadline is approaches. Even in the absence of a clear deadline some
priorities may incease in time \cite{blanchard}, others may decrease.
Sometimes external factors change suddenly a task's priority-- for
example, the priority of watering the lawn suddenly diminishes if it
starts raining.  The possibility of dropping tasks, either by not allowing
them on the queue, or by simply deleting them from the queue, could also
affect the waiting time distributions. Tasks could be dropped if they were
not executed for a considerable time interval, and thus become irrelevant,
or when the individual is very busy, or some may be simply forgotten.
Obviously, the precise impact on the waiting time distribution will depend
on the implementation of the task dropping conditions. It is important to
understand if any or all of these processes could change the universality
class of the waiting time distribution.

{\it Model limitations:} The studied datasets do not capture all tasks an
individual is involved in, but only the timing of selected activities,
like sending emails or borrowing books from the library.  Yet, we must
consider the fact that between any two recorded events individuals
participate in many other non-recorded activities. For example, if we find
that an individual clicks on a new document every few seconds, likely
he/she is fully concentrating on web browsing. However, when we notice a
break of hours or days between two consecutive clicks, it is clear that in
the meantime the individual was involved in a myriad of other activities
that were not visible to us. The queuing models discussed here were
designed to take into consideration all human activities, as we assume
that the priority list of an individual contains all tasks the person is
involved in. Currently an understanding of the interplay between the {\it
recorded} and the {\it invisible} activities is still lacking.

{\it Task optimization:} The order in which we execute different tasks is
often driven by optimization: we try to minimize the total time, or some
cost functions. This is particularly relevant if the execution times
depend on the order in which the tasks are executed. For example, often
executing a certain task might be faster if we execute some other
preparatory tasks before, and not in the inverse order. In principle
optimization could be incorporated into the studied models by assuming
that they determine the priority of the tasks. Optimization raises several
important questions for future work: How should we model optimization
driven queueing processes? Can they also lead to power laws, and if so,
will they result in new universality classes?

{\it Correlations:} So far we have focused on the origin of the various
distributions observed in human dynamics.  Distributions offer little
information, however, about potential correlations present in the observed
time series. Such correlations were documented in Ref. \cite{Maya},
observing that the correlation function of the interevent time series for
printing job arrivals decayed as a power law. Are such temporal
correlations present in other systems as well? What is their origin? Can
the queuing models predict such correations? Answers to these questions
could not only help better understand human dynamics, but could also aid
in distingushing the various models from each other.

{\it Network effects:} In seaching for the explanation for the observed
heavy tailed human activity patterns we limited our study to the
properties of single queues. In reality none of our actions are perforned
independently--most of our daily activity is embedded in a web of actions
of other individuals \cite{stanley05,holme05}. Indeed, the timing of an
email sent to user A may depend on the time we receive an email from user
B. An important future goal is to understand how the various human
activities and their timing is affected by the fact that the individuals
are embedded in a network environment.

{\it Non-human activity patterns:} Heavy tailed interevent time
distributions do not occur only in human activity, but emerge in many
natural and technological systems. For example, Omori's law on earthquaqes
\cite{omori,apostol} records heavy tailed interevent times between
consecutive seismic activities; measurements indicate that the fishing
patters of seabirds also display heavy tailed statistics
\cite{Viswanathan}; plasticity paterns \cite{zapperi} and avalanches in
lungs \cite{suki} show similar power law interevent times. While a series
of models have been proposed to capture some of these processes
individually, there is also a possibility that some of these modeling
frameworks can be reduced to various queuing processes. Some of the
studied queuing models show close relationship to several models designed
to capture self-organized criticality
\cite{Sneppen,Jensen,Bak,Maya1,Flyvbjerg1,Boer}. Could the mechanisms be
similar at some fundamental level? Even if such higher degree of
universality is absent, understanding the mechanisms and queuing processes
that drive human dynamics could help us better understand other natural
phenomena as well, from the timing of chemical reactions in a cell to the
temporal records of major economic events \cite{stanley05} or the timing
of events in manufacturing processes and supply chains
\cite{helbing1,helbing2,helbing3} or panic \cite{helbing}.

\appendix

\section{Preferential selection protocol}
\label{sec:preferential}

As we discussed in Sect. \ref{sec:discussion}, one possible modification
of the priority model introduced and studied in Sect. \ref{sec:barabasi}
involves the assumption that we do not always choose the highest priority
task for execution, but rather the tasks are chosen stochastically, in
increasing function of their priority.  That is, the probability to choose
a task with priority $x$ for execution in a unit time is $\Pi(x) \sim
x^\gamma$, where $\gamma$ is a predefined parameter of the model. This
parameter allows us to interpolate between the random choice limit (ii)
($\gamma=0$, $p=0$) and the deterministic case, when always the highest
priority item is chosen for execution (iii) ($\gamma=\infty$, $p=1$). Note
that this parameterization captures the scaling of the model discussed in
Sect.  \ref{sec:barabasi} only in the $p \to 0$ and $p\to 1$ limits, but
not for intermediate $p$ values. That is, the two limits of this model map
into extreme limits of the model introduced in Sect. \ref{sec:barabasi},
but the intermediate $p$ and $\gamma$ values do not map into each other.

The probability that a task with priority $x$ waits a time interval $t$
before execution is ${f }(x,t)=(1-\Pi(x))^{t-1}\Pi(x)$. The average
waiting time of a task with priority $x$ is obtained by averaging over $t$
weighted with ${f} (x,t)$, providing

\begin{equation}
t_w(x)=\sum_{t=1}^\infty t f(x,t) =\frac{1}{\Pi(x)}\sim \frac{1}
{x^\gamma}, \label{tx}
\end{equation}

\noindent {\it i.e.} the higher an item's priority, the shorter is
the average time it waits before execution. To calculate $P(\tau)$
we use the fact that the priorities are chosen from the $\eta(x)$
distribution, i.e. $\eta(x)dx=P(\tau) d\tau$, which gives

\begin{equation}
w(t_w) \sim \frac{\eta(t_w^{-1/\gamma})}{t_w^{1+1/\gamma}}\ ,
\label{Pt}
\end{equation}

\noindent providing the relationship (\ref{alphagamma}) between $\alpha$
and $\gamma$, and indicating that with changing $\gamma$ we can
continuously tune $\alpha$ as well. In the $\gamma\to\infty$ limit, which
converges to the strictly priority based deterministic choice ($p=1$) in
the model, Eq. (\ref{Pt}) predicts $w(t_w) \sim t_w^{-1}$, in agreement
with the numerical results (Fig 3a), as well as the empirical data on the
email interarrival times (Fig 2a). In the $\gamma=0$ ($p=0$) limit
$t_w(x)$ is independent of $x$, thus $w(t_w)$ converges to an exponential
distribution, as shown in Fig. 3b.

The apparent dependence of $w(t_w)$ on the $\eta(x)$ distribution
from which the agent chooses the priorities may appear to represent
a potential problem, as assigning priorities is a subjective
process, each individual being characterized by its own $\eta(x)$
distribution. According to Eq. (\ref{Pt}), however, in the $\gamma
\to \infty$ limit $w(t_w)$ is independent of $\eta(x)$. Indeed, in
the deterministic limit the uniform $\eta(x)$ can be transformed
into an arbitrary $\eta'(x)$ with a parameter change, without
altering the order in which the tasks are executed
\cite{queue-cohen}. This insensitivity of the tail to $\eta(x)$
explains why, despite the diversity of human actions, encompassing
both professional and personal priorities, most decision driven
processes develop a heavy tail.

\section{Exact solution of the priority queue model with $L=2$}
\label{app:l2}

Consider the model discussed in Sect. \ref{sec:barabasi} \cite{barabasi05}
with $L=2$ \cite{vazquez05}. The task that has been just selected and its
priority has been reassigned will be called the new task, while the other
task will be called the old task. Let $\eta(x)$ and
$R(x)=\int_0^xdx\tilde{\eta}(x)$ be the priority probability density
function (pdf) and distribution function of the new tasks, which are
given. In turn, let $\tilde{\eta}(x,t)$ and
$\tilde{R}(x,t)=\int_0^xdx\tilde{\eta}(x,t)$ be the priority pdf and
distribution function of the old task in the $t$-th step. At the
$(t+1)$-th step there are two tasks on the list, their priorities being
distributed according to $R(x)$ and $\tilde{R}(x,t)$, respectively. After
selecting one task the old task will have the distribution function

\begin{equation}
\tilde{R}(x,t+1) = \int_0^x dx^\prime \tilde{\eta}(x^\prime,t) q(x^\prime)
+ \int_0^{x} dx^\prime \eta(x) \tilde{q}(x^\prime,t)\ ,
\label{RtRt}
\end{equation}

\noindent where

\begin{equation}
q(x) = p[1-R(x)] +(1-p)\frac{1}{2}
\label{q}
\end{equation}

\noindent is the probability that the new task is selected given
the old task has priority $x$, and

\begin{equation}
\tilde{q}(x) = p[1-\tilde{R}(x,t)] +(1-p)\frac{1}{2}
\label{q1}
\end{equation}

\noindent is the probability that the old task is selected given the new
task has priority $x$. In the stationary state,
$\tilde{R}(x,t+1)=\tilde{R}(x,t)$, thus from (\ref{RtRt}) we obtain

\begin{equation}
\tilde{R}(x) = \frac{1+p}{2p} \left[ 1 -
\frac{1}{1+\frac{2p}{1-p}R(x)} \right]\ .
\label{Rx}
\end{equation}

Next we turn our attention to the waiting time distribution.  Consider a
task with priority $x$ that has just been added to the queue. The
selection of this task is independent from one step to the other.
Therefore, the probability that it waits $\tau_w$ steps is given by the
product of the probability that it is not selected in the first $\tau_w-1$
steps and that it is selected in the $\tau_w$-th step.  The probability 
that
it is not selected in the first step is $\tilde{q}(x)$, while the
probability that it is not selected in the subsequent steps is $q(x)$.
Integrating over the new task's possible priorities we obtain

\begin{equation}
P(\tau_w) = \left\{
\begin{array}{ll}
\int_0^\infty dR(x) \left[ 1 - \tilde{q}(x) \right] \ , & \tau_w=1\\
\\
\int_0^\infty dR(x) \tilde{q}(x) \left[ 1 - q(x) \right]
q(x)^{\tau_w-2}\ , & \tau_w>1\\
\end{array}
\right.
\label{ptau0}
\end{equation}

\noindent Using (\ref{q})-(\ref{Rx}) and integrating (\ref{ptau0}) we 
finally obtain

\begin{equation}
P(\tau_w) = \left\{
\begin{array}{ll}
1 - \frac{1-p^2}{4p} \ln \frac{1+p}{1-p}\ , & \tau_w=1\\
\\
\frac{1-p^2}{4p(\tau_w-1)} \left[ \left(\frac{1+p}{2}\right)^{\tau_w-1}
- \left(\frac{1-p}{2}\right)^{\tau_w-1} \right]\ , & 
\tau_w>1
\end{array}
\right.
\label{ptau2}
\end{equation}

\noindent Note that $P(\tau_w)$ is independent of the $\eta(x)$ pdf from
which the tasks are selected. Indeed, what matters for task selection is
their relative order with respect to other tasks, resulting that all
dependences in (\ref{q})-(\ref{Rx}) and (\ref{ptau0}) appears via $R(x)$.

\section{The asymptotic characteristics of $P(\tau_w)$}
\label{general}

In Sect. \ref{sec:barabasi} we focused on a model with fixed queue length
$L$, demonstrating that it belongs to a new universality class with
$\alpha=1$. Next we derive a series of results that apply to any queuing
model that has a {\it finite queue length}, and is characterized by an
{\it arbitrary task selection protocol} \cite{vazquez05}. In each time
step there are $L$ tasks in the queue and one of them is executed.
Therefore

\begin{equation}
\sum_{i=1}^t\tau_i + \sum_{i=1}^{L-1}\tau_i^\prime = Lt\ ,
\label{tauLt}
\end{equation}

\noindent where $\tau_i$ is the waiting time of the task executed at the
$i$-th step and $\tau_i^\prime$, $i=1,\ldots,L-1$, is the time interval
that task $i$, that is still active at the $t$-th step, has already spent
on the queue. The first term in the l.h.s. of (\ref{tauLt}) corresponds to
the sum of the waiting times experienced by the $t$ tasks that were
executed in the $t$ steps since the beginning of the queue, while the
second term describes the sum of the waiting times of the $L-1$ tasks that
are still on the queue after the $t$ step. Given that in each time step
each of the $L$ tasks experience one time step delay, the sum on the
l.h.s. should equal $Lt$. From (\ref{tauLt}) it follows that

\begin{equation}
\langle \tau_w\rangle \equiv \lim_{t\rightarrow\infty}
\frac{1}{t}\sum_{i=1}^t\tau_i = L -
\lim_{t\rightarrow\infty}\frac{1}{t}\sum_{i=1}^{L-1}\tau_i^\prime\ .
\label{tauave}
\end{equation}

\noindent If all active tasks have a chance to be executed sooner or
later, like the case for the model studied in Sects.  \ref{sec:barabasi}
in the $0\leq p<1$ regime \cite{barabasi05}, we have
$\langle\tau_w^\prime\rangle\leq\langle \tau_w\rangle$ and the last term 
in
(\ref{tauave}) vanishes when $t\rightarrow\infty$. In contrast, for $p=1$
the numerical simulations \cite{barabasi05} indicate that after some
transient time the most recently added task is always executed, while
$L-1$ tasks remain indefinitely in the queue. In this case
$\tau_i^\prime\sim t$ in the $t\rightarrow\infty$ limit and the last term
in (\ref{tauave}) is of the order of $L-1$. Based on these arguments we
conjecture that the average waiting time of executed tasks is given by

\begin{equation}
\langle \tau_w\rangle = \left\{
\begin{array}{ll}
L \ , & 0\leq p<1\\
1\ , & p=1\ ,
\end{array}
\right.
\label{tauaveL}
\end{equation}

\noindent which is corroborated by numerical simulations (see Fig.
\ref{fig:l2}b).

It is important to note that the equality in (\ref{tauave}) is independent
of the selection protocol, allowing us to reach conclusions that apply
beyond the model discussed in Sect.  \ref{sec:barabasi}. From
(\ref{tauave}) we obtain

\begin{equation}
\langle\tau_w\rangle \leq L\ .
\label{tauaveLineq}
\end{equation}

\noindent From this constraint follows that $P(\tau_w)$ must decay
faster than $\tau_w^{-2}$ when $\tau_w\rightarrow\infty$, otherwise
$\langle \tau_w \rangle$ would not be bounded. Indeed, it is easy to
see that for any $\alpha<2$ the average waiting time $\langle \tau_w
\rangle$ diverges for Eq. (\ref{E:eq1}). Thus, when
$\tau_w\rightarrow\infty$, we must either have

\begin{equation}
P(\tau_w) \sim a\tau_w^{-\alpha}\ ,\alpha>2
\label{ptau1}
\end{equation}

\noindent or

\begin{equation}
P(\tau_w) = \tau_w^{-\alpha} f \left( \frac{\tau_w}{\tau_0} \right)\ ,
\label{ptauscaling}
\end{equation}

\noindent where $\tau_0>0$ and $f(x)={\cal O}(bx^{\alpha-2})$ when
$x\rightarrow\infty$, where $b$ is a constant. That is, each time an
$\alpha<2$ exponent is observed (as it is for the empirical data discussed
in Sect. \ref{sec:empirical}), an exponential cutoff must accompany the
scaling. For example, for the model discussed above with $L=2$ and $0\leq
p<1$ we have $\alpha=1$ and $f(x)$ decays exponentially (\ref{ptau00}), in
line with the constraint discussed above.

\section{Transitions between the two universality classes}
\label{sec:transition}

A basic difference between the models discussed in Sect. \ref{sec:cobham}
and Sects. \ref{sec:barabasi} is the capacity of the queue. Our results
indicate that the model without limitation on the queue length displays
$\alpha=3/2$, rooted in the fluctuations of the queue length. In contrast,
the model with fixed queue length (Sect. \ref{sec:barabasi}) has
$\alpha=1$, rooted in the queuing of the low priority tasks on the
priority list. If indeed the limitation in the queue length plays an
important role, we should be able to develop a model that can display a
transition from the $\alpha=3/2$ to the $\alpha=1$ universality class as
we limit the fluctuations in the queue length. In this section we study
such a model, interpolating between the two observed scaling regimes. We
start from the model discussed in Sect.  \ref{sec:cobham}, and impose on
it a maximum queue length $L$. This can be achieved by altering the
arrival rate of the tasks: when there are $L$ tasks in the queue no new
tasks will be accepted until at least one of the tasks is executed.  
Mathematically this implies that the arrival rate depends on the queue
length as

\begin{equation}
\lambda_\ell=\left\{
\begin{array}{ll}
\lambda\ , & 0\leq \ell<L\\
0\ , & \ell=L\ .
\end{array}
\right.
\label{lambdal}
\end{equation}

\noindent In the stationary state the queue length distribution
$P(\ell)$ satisfies the balance equation

\begin{equation}
\lambda_{\ell-1} P(\ell-1) + \mu_{\ell+1} P(\ell+1) =
\left( \lambda_\ell + \mu_\ell \right) P(\ell)\ ,
\label{Llbalance}
\end{equation}

\noindent where

\begin{equation}
\mu_\ell=\left\{
\begin{array}{ll}
0\ , & \ell=0\\
\mu\ , & 0<\ell\leq L\ .
\end{array}
\right.
\label{mul}
\end{equation}

\noindent From (\ref{Llbalance}) we obtain the queue length
distribution as

\begin{equation}
P(\ell) = \frac{1-\rho}{1-\rho^{L+1}} \rho^\ell\ ,
\label{LL}
\end{equation}

\noindent suggesting the existence of three scaling regions.

\begin{figure}

\centerline{\includegraphics[width=3.2in]{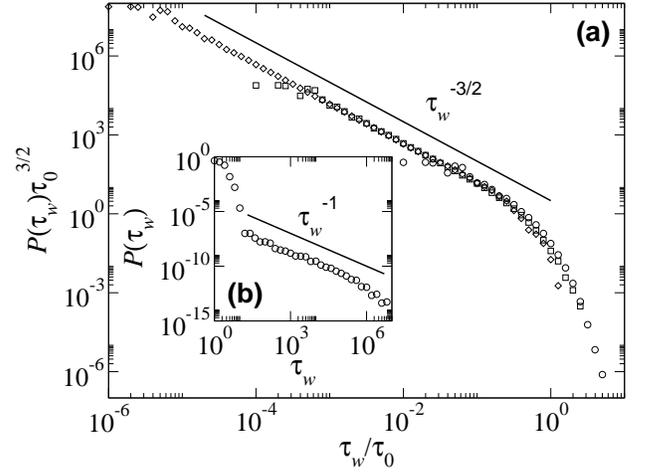}}

\caption{Waiting time distribution for tasks in the queueing model
discussed in Sect. \ref{sec:transition}, with a maximum queue length $L$.  
The waiting time distribution is plotted for three $L$ values: $L=10$
(circles), $L=100$ (squares) and $L=1000$ (diamonds). The data has been
rescaled to emphasize the scaling behavior
$P(\tau_w)=\tau_w^{-3/2}f(\tau_w/\tau_0)$, where $\tau_0\sim L^2$. In the
inset we plot the waiting time for $\rho=10^6$, showing the crossover to
the model discussed in Sect.  \ref{sec:barabasi} in the limit
$\rho\rightarrow\infty$ and $L$ fixed.}

\label{F:fig6}
\end{figure}

{\it Subcritical regime}, $\rho \ll 1$: If the arrival rate of the tasks
is much smaller than the execution rate, the fact that the queue length
has an upper bound has little significance, since $\ell$ will rarely reach
its upper bound $L$, but will fluctuate in the vicinity of $\ell=0$. This
regime can be reached either for $\rho\ll1$ and $L$ fixed or for $\rho<1$
and $L\gg1$. Therefore, in this case the waiting time distribution is well
approximated by that of the model with an unlimited queue length,
displaying the scaling predicted by Eq. (\ref{w0}), {\it i.e.} either
exponential, or a power law with $\alpha=3/2$, coupled with an exponential
cutoff (see Fig. \ref{F:fig6}a).

{\it Critical regime}: For $\rho=1$ we observe an interesting interplay
between the queue length and $L$. Normally in this critical regime
$\ell(t)$ should follow a random walk with the return time probability
density scaling with exponent $3/2$. However, the limitation imposed on
the queue length limits the power law waiting time distribution predicted
by Eq. (\ref{w0}), introducing a cutoff (see Fig. \ref{F:fig6}b). Indeed
having the number of tasks in the queue limited allows each task to be
executed in a finite time.

{\it Supercritical regime}: When $\rho\gg1$ from (\ref{LL}) follows
that

\begin{equation}
{\cal L}_\ell = \left\{
\begin{array}{ll}
{\cal O}(\rho^{-1})\ , & 0\leq \ell<L\\
1 - {\cal O}(\rho^{-1})\ , & \ell=L\ ,
\end{array}\right.
\label{LL1}
\end{equation}

\noindent {\it i.e.} with probability almost one the queue is filled. 
Thus, in the supercritical regime $\rho\gg1$ new tasks are added to the 
queue immediately after a task is executed. If we take the number of 
executed tasks as a new reference time then this model corresponds to the 
one discussed in Sect. \ref{sec:barabasi}, displaying $\alpha=1$ 
\cite{barabasi05}, as supported by the numerical simulations (see Fig. 
\ref{F:fig6}a).

\section{Heavy tailed input distributions}
\label{app:service}

In this Appendix we study the model discussed in Sec. \ref{sec:cobham}
with a heavy tailed service time distribution $h(s)\sim s^{-\beta}$ with
$0<\beta<1$. In this case it has been shown that \cite{abate97}

\begin{equation}
P(\tau_w)\sim \tau_w^{-\beta}\ . \label{wbeta}
\end{equation}

\noindent This result is a consequence of the generalized limit
theorem for heavy tailed distributions \cite{fellerII}. Let us focus
on a selected task and assume that $m$ tasks need to be executed
before it. Therefore, the selected task's waiting time is given by

\begin{equation}
\tau_w = \sum_{l=1}^m s_l\ , \label{twm}
\end{equation}

\noindent where $s_l$ is the service time of the $l$-th task
executed before the given task. Equation (\ref{twm}) represents the
sum of $m$ independent and identically distributed random variables,
with pdf $h(s)\sim s^{-\beta}$, which is known to follow a pdf with
the same heavy tail, and thus resulting in (\ref{wbeta}). Hence, in
this case the heavy tail in the waiting time distribution is a
consequence of the heavy tails in the service time distribution.

\begin{acknowledgments}

We wish to thank Diana Kormos Buchwald and Tilman Sauer for providing us
the dataset capturing the Einstein correspondence, and for helping us
understand many aspects of Einstein's life and correspondence patterns.
Similarly, we wish to thank Alison M Pearn from the Darwin Correspondence
Project for providing us the record of Darwin's communications and
Thurston Miller for providing us the library visit dataset. We have
benefited from useful discussions with L.A.N. Amaral and D. Stouffer.  
A.-L. B., A. V. and Z. D. are supported by NSF ITR 0426737, NSF ACT/SGER
0441089 awards and by an award from the James S. McDonnell Foundation.
J.G.O. is supported by FCT (Portugal) grant No. SFRH/BD/14168/2003. I.K. 
wish to thank B. Jank\'o and the Institute for Theoretical Sciences at 
University of Notre Dame for their hospitality during this collaboration, 
as well as the National Office of Research and Technology in Hungary for 
support.

\end{acknowledgments}


\begin{thebibliography}{99}

\bibitem{fellerII} W. Feller, {\it An introduction to probability theory 
and its applications} Vol. 2 (Wiley, 2nd Edition, New York, 1971).

\bibitem{Haight67} F. A. Haight, {\it Handbook of the Poisson
Distribution} (Wiley, New York, 1967).

\bibitem{erlang} A. K. Erlang, Nyt. Tidsskrift for Matematik B {\bf 20} 
1909; Elektrotkeknikeren {\bf 13} (1917).

\bibitem{phone-design} H. R. Anderson, {\it Fixed Broadband Wireless
System Design} (Wiley, New York, 2003).

\bibitem{barabasi05} A.-L. Barab\'asi, Nature {\bf 207}, 435 (2005).

\bibitem{joao05} J. G. Oliveira and A.-L. Barab\'asi, Nature (in press).

\bibitem{zoli} Z. Dezs\H{o}, E. Almaas, A. Luk\'acs, B. R\'acz, I.  
Szakad\'at and A.-L. Barab\'asi. physics/0505087.

\bibitem{vazquez05} A. V\'azquez, physics/0506126.

\bibitem{stanley} H. E. Stanley, {\it Introduction to phase transitions 
and critical phenomena} (Oxford Univ. Press, Oxford, 1987).

\bibitem{review1} R. Albert and A.-L. Barab\'asi, Rev. Mod. Phys. {\bf 
74}, 47 (2002).

\bibitem{review2} S. N. Dorogovtsev and J. F. F. Mendes, {\it Evolution 
of Networks: From biological nets to the Internet and the WWW}, Oxford 
Univ. Press, Oxford, 2003).

\bibitem{vespignani-book} R. Pastor-Satorras and A. Vespignani, {\it 
Evolution and structure of the Internet: A Statistical Physics Approach} 
(Cambridge Univ. Press, Cambridge, 2004).

\bibitem{poisson} S.-D. Poisson, {\it Recherches sur la probabilit\'e des
jugements en mati\`ere criminelle et en mati\`ere civile, pr\'ec\'ed\'ees
des r\`egles g\'en\'erales du calcul des probabilit\'es.} (Bachelier,
Paris, 1837).

\bibitem{call-center} P. Reynolds, {\it Call Center Staffing} (The Call
Center School Press, Lebanon, TN, 2003).

\bibitem{inventory} J. H. Greene, {\it Production and Inventory Control
Handbook} (McGraw-Hill, New York, 3 ed, 1997).

\bibitem{Instant} C. Dewes, A. Wichmann, and A. Feldman, An Analysis of
Internet Chat Systems, {\it Proc. 2003 ACM SIGCOMM Conf. on Internet
Measurement (IMC-03)}, ACM Press, New York, 2003).

\bibitem{supercomputers} S. D. Kleban and S. H. Clearwater, Hierarchical
Dynamics, Interarrival Times and Performance, {\it Proc. of SC'03},
November 15-21, 2003, Phonenix, AZ, USA.

\bibitem{ftp} V. Paxson and S. Floyd, Wide-Area Traffic: The Failure of
Poisson Modeling, {\it IEEE/ACM Tansactions in Networking} {\bf 3}, 226
(1995).

\bibitem{mainardi00} F. Mainardi, M. Raberto, R. Gorenflo and E. 
Scalas, cond-mat/0006454

\bibitem{mainardi02} M. Raberto, E. Scalas, F. Mainardi, cond-mat/0203596

\bibitem{economic1} V. Plerou, P. Gopikirshnan, L. A. N. Amara, X.
Gabaix, and H. E. Stanley, Phys. Rev. E {\bf 62}, R3023 (2000).

\bibitem{economic2} J. Masoliver, M. Montero and G. H. Weiss, Phys. Rev. E
{\bf 67}, 021112 (2003).

\bibitem{games} T. Henderson and S. Nhatti, Modelling user behavior in
networked games, {\it Proc. ACM Multimedia 2001}, Ottawa, Canada,
212--220, 30 September--5 October (2001).

\bibitem{eckmann} J.-P. Eckmann, E. Moses and D. Sergi, Proc. Nat.
Acad. Sci. USA {\bf 101}, 14333 (2004).

\bibitem{ebel} H. Ebel, L.-I. Mielsch, and S. Bornholdt, Phys. Rev. E {\bf
66}, R35103 (2002).

\bibitem{Maya} U. Harder and M. Paczuski, cs.PF/0412027.

\bibitem{racz04} B. R\'acz and A. Luk\'acs, High density compression of 
log files, Data compression conference, IEEE Computer Society Press 
(2004).

\bibitem{einstein} {\it The collected papers of Albert Einstein}, Vol. 
{\bf 1,5,8,9} (Princeton Univ. Press, 1993-2004).

\bibitem{darwin} {\it The correspondence of Charles Darwin}, Vol. {\bf 
1-14} (Cambridge Univ. Press, Cambridge, 1984-2004).

\bibitem{freud} Freud Museum, London, http://www.freud.org.uk/.

\bibitem{queue-cohen} J. W. Cohen, {\it The Single Server Queue} (North
Holland, Amsterdam, 1969).

\bibitem{gross98} D. Gross and C. M. Harris, Fundamentals of queueing 
theory (Wiley, New York, 1998).

\bibitem{Crovella} M. Crovella and Bestavros, IEEE/ACM Trans.
Netw. {\bf 5}, 835 (1997).

\bibitem{Mitzenmacher} M. Mitzenmacher, Internet Mathematics {\bf 1},
226-251 (2004).

\bibitem{Miller} G. A. Miller, Psychological Review, {\bf 63}, 8197
(1956).

\bibitem{baddeley94} A. Baddeley, Psychological Bulletin, {\bf 101}, 353 
(1994); {\it ibid} {\bf 101}, 668 (1994).

\bibitem{Cobham} A. Cobham, J. Oper. Res. Soc. Amer. {\bf 2}, 70 (1954).

\bibitem{abate97} J. Abate, Queueing Systems {\bf 25}, 173 (1997).

\bibitem{redner01} S. Redner, {\it A guide to first-passage processes}
(Cambridge University Press, New York, 2001).

\bibitem{havlinRW} {\it Diffusion and reactions in fractals and disordered
systems}, Editors D. Ben-Avraham and S. Havlin (Cambridge University
Press, Cambridge, 2000).

\bibitem{jeong03} H. Jeong, Z. N\'eda, and A.-L. Barab\'asi, Europhy.  
Lett.  {\bf 61}, 567 (2003).

\bibitem{barabasi1} A.-L. Barab\'asi and R. Albert, Science {\bf 
286}, 509 (1999).

\bibitem{barabasi3} A.-L. Barab\'asi, H. Jeong, R. Ravasz, Z. N\'eda, T. 
Vicsek, and A. Schubert, Physica A {\bf 311}, 590 (2002).

\bibitem{internet1} R. Pastor-Satorras, A. V\'azquez and A. Vespignani, 
Phys. Rev. Lett. {\bf 87}, 258701 (2001).

\bibitem{internet2} A. V\'azquez, R. Pastor-Satorras, and A. Vespignani, 
Phys. Rev. E {\bf 65}, 0666130 (2002).

\bibitem{blanchard} Ph. Blanchard and M. O. Hongler, preprint.

\bibitem{stanley05} K. Yamasaki, L. Muchnik, S. Havlin, A. Bunde, and H. 
E. Stanley, Proc. Nat. Acad. Sci. {\bf 102}, 9424 (2005).

\bibitem{holme05} P. Holme, Phys. Rev. E {\bf 71}, 046119 (2005).

\bibitem{omori} F. J. Omori, Coll. Sci. Imper. Univ. Tokyo {\bf 7}, 111
(1894).

\bibitem{apostol} B.-F. Apostol, Euler's transform and a generalized 
Omori's law, prewprint.

\bibitem{Viswanathan} G. M. Viswanathan {\it et al}, Nature {\bf 381}, 413
(1996).

\bibitem{zapperi} S. Zapperi, A. Vespignani and H. E. Stanley
Nature {\bf 388}, 658 (1997).

\bibitem{suki} B. Suki, A.-L. Barab\'asi, Z. Hantos, F. Petak, and H. E.
Stanley, Nature {\bf 368}, 615 (1994).


\bibitem{Sneppen} P. Bak and K. Sneppen, Phys. Rev. Lett. {\bf 71}, 4083
(1993).

\bibitem{Jensen} H. J. Jensen, {\it Self-Organised Criticality}
(Cambridge Univerity Press, Cambridge, 1998).

\bibitem{Bak} P. Bak, C. Tang, and K. Wiesenfeld, Phys. Rev. Lett. {\bf
59}, 381 (1987).

\bibitem{Maya1} M. Paczuski, S. Maslov, and P. Bak, Phys. Rev. E
{\bf 53}, 414 (1996).

\bibitem{Flyvbjerg1} H. Flyvbjerg, K. Sneppen, and P. Bak, Phys. Rev.
Lett. {\bf 71}, 4087 (1993).


\bibitem{Boer} J. de Boer, B. Derrida, H. Flyvbjerg, A.D. Jackson,
and T. Wettig, Phys. Rev. Lett. {\bf 73} 906- (1994).

\bibitem{helbing1} B. A. Huberman and D. Helbing, Europhys. Lett. {\bf 
47}, 196 (1999).

\bibitem{helbing2} D. Helbing, S. Lammer, T. Seidel, {\it et al}, Phys. 
Rev. E {\bf 70}, 066116 (2004).

\bibitem{helbing3} D. Helbing, S. Lammer, U. Witt, {\it et al}, Phys. 
Rev. E {\bf 70}, 056118 (2004).

\bibitem{helbing} D. Helbing, I. Farkas and T. Vicsek, Nature {\bf 407},
487 (2000).

\end{thebibliography}
\end{document}